\documentclass[journal=jacsat,manuscript=article]{achemso}
\setkeys{acs}{keywords = true}
\usepackage[version=3]{mhchem}
\usepackage{hyphenat}
\usepackage{array, caption}%
\mciteErrorOnUnknownfalse
\restylefloat*{table}

\author{Kimmo Mustonen}
\affiliation{University of Vienna, Faculty of Physics, 1090 Vienna, Austria}
\email{kimmo.mustonen@univie.ac.at}
\altaffiliation{Contributed equally to this work}

\author{Christoph Hofer}
\affiliation{Eberhard Karls University of Tuebingen, Institute for Applied Physics, 72076 Tuebingen, Germany}
\alsoaffiliation{NMI Natural and Medical Sciences Institute at the University of Tuebingen, Markwiesenstr. 55, D-72770 Reutlingen, Germany}
\alsoaffiliation{University of Antwerp, EMAT, 2020 Antwerp, Belgium}
\altaffiliation{Contributed equally to this work}

\author{Peter Kotrusz}
\affiliation{Danubia NanoTech s.r.o., Bratislava}
\alsoaffiliation{Institute of Electrical Engineering, SAS, Bratislava}

\author{Alexander Markevich}
\affiliation{University of Vienna, Faculty of Physics, 1090 Vienna, Austria}

\author{Martin Hulman}
\affiliation{Danubia NanoTech s.r.o., Bratislava}
\alsoaffiliation{Institute of Electrical Engineering, SAS, Bratislava}

\author{Clemens Mangler}
\affiliation{University of Vienna, Faculty of Physics, 1090 Vienna, Austria}

\author{Toma Susi}
\affiliation{University of Vienna, Faculty of Physics, 1090 Vienna, Austria}

\author{Timothy J. Pennycook}
\affiliation{University of Antwerp, EMAT, 2020 Antwerp, Belgium}

\author{Karol Hricovini}
\affiliation{CY Cergy Paris University, LPMS, 95 031 Cergy-Pontoise, France}

\author{Christine M. Richter}
\affiliation{CY Cergy Paris University, LPMS, 95 031 Cergy-Pontoise, France}

\author{Jannik C. Meyer}
\affiliation{Eberhard Karls University of Tuebingen, Institute for Applied Physics, 72076 Tuebingen, Germany}
\alsoaffiliation{NMI Natural and Medical Sciences Institute at the University of Tuebingen, Markwiesenstr. 55, D-72770 Reutlingen, Germany}

\author{Jani Kotakoski}
\affiliation{University of Vienna, Faculty of Physics, 1090 Vienna, Austria}
\email{jani.kotakoski@univie.ac.at}

\author{Viera Sk\'akalov\'a}
\affiliation{University of Vienna, Faculty of Physics, 1090 Vienna, Austria}
\alsoaffiliation{Danubia NanoTech s.r.o., Bratislava}
\alsoaffiliation{Institute of Electrical Engineering, SAS, Bratislava}
\email{skakalova@danubiananotech.com}

\title[An \textsf{achemso} demo]
{Towards Exotic Layered Materials: 2D Cuprous Iodide}


\makeatletter
\newcommand*{\forcekeywords}{
  \acs@keywords@print
  \let\acs@keywords@print\relax
}
\makeatother

\begin{document}
\forcekeywords




\newpage

\begin{abstract}

Heterostructures composed of two\hyp{}dimensional (2D) materials are already opening many new possibilities in such fields of technology as electronics and magnonics, but far more could be achieved if the number and diversity of 2D materials is increased. So far, only a few dozen 2D crystals have been extracted from materials that exhibit a layered phase in ambient conditions, omitting entirely the large number of layered materials that may exist in other temperatures and pressures. Here, we demonstrate how these structures can be stabilized in 2D van der Waals stacks under room temperature via growing them directly in graphene encapsulation by using graphene oxide as the template material. Specifically, we produce an ambient stable 2D structure of copper and iodine, a material that normally only occurs in layered form at elevated temperatures between 645 and 675~K. Our results establish a simple route to the production of more exotic phases of materials that would otherwise be difficult or impossible to stabilize for experiments in ambient.

\end{abstract}


\section{Main}

Current 2D materials largely derive from van der Waals (vdW)\hyp{}layered bulk structures. However, only a limited number of such structures exist under ambient conditions and in total, only a few dozen 2D crystals have been successfully synthesized or exfoliated. While the unusual properties of graphene make it an interesting object of investigation itself~\cite{novoselov2004electric}, it can also serve as a substrate to stabilize other, less obvious 2D materials. These include materials that do not by themselves form 2D phases, such as the covalent SiO$_2$~\cite{huang2012direct}, pseudo\hyp{}ionic PbI$_2$~\cite{sinha2020atomic}, and metallic CuAu~\cite{zagler2020cuau}.

In the same spirit, layers of graphene have also been used to encapsulate materials. Metal atoms (in some cases forming nitrides \cite{al2016two, pecz2021indium}) have been intercalated between a monocrystalline SiC surface and graphene to produce 2D metamaterials \cite{rosenzweig2020large, forti2020semiconductor, briggs2020atomically}. In other studies the encapsulation strategy has been applied in {\it in situ} transmission electron microscopy (TEM) observations of dynamics in liquids~\cite{kuhne2018reversible,kelly2018nanometer} and for protection of electron beam sensitive materials~\cite{zan2013control, algara2013pristine}. In addition, the inert and impermeable graphene envelope can also stabilize 2D layers of weakly bound molecules and atoms, and islands of C$_{60}$ fullerenes~\cite{mirzayev2017buckyball} and noble gases~\cite{langle20202d} have been observed in graphene encapsulation. In the latter two examples especially, the significant, over 1~GPa (or 10$^4$~atm) pressure associated with the vdW forces between graphene layers~\cite{vasu2016van}, is crucial to constraining the degrees of freedom and compelling the encapsulated species to assume and retain the 2D crystalline phase. The concept of encapsulation however is not limited to mere vdW materials and indeed, it entails a much greater variety of similarly constrained 2D structures that in their bulk form exhibit a vdW-layered phase only either at elevated temperatures or pressures. One of such material among countless that have been predicted~\cite{mounet2018two} is the $\beta$-phase of CuI that is only stable at a narrow temperature range of 645-675 K and therefore \cite{sakuma1988crystal, keen1994determination}, exfoliation of monolayers from such crystals would be rather complicated. The 3D $\gamma$-phase of CuI, in contrast, is stable at ambient conditions and has been known for a high Seebeck coefficient (thermoelectrics) and appreciable optoelectric properties. Extremely thin layers of this non-layered cubic $\gamma$-phase of CuI have been successfully prepared and combined with 2D WS$_{2}$ and WSe$_{2}$ \cite{yao2018synthesis}. Nevertheless, the truly 2D phase of CuI, to our knowledge, has not yet been prepared and its properties remain unexplored. In fact, two distinct configurations of the 2D CuI structure were recently predicted by Mounet \textit{et al.} \cite{mounet2018two}, both unstable at ambient conditions; our primary motivation to this work therefore was to find out whether and which of the predicted structures could be synthesized.

Here, we demonstrate a method for the stabilization of a single layer of the high temperature vdW-layered $\beta$-CuI at ambient conditions by using graphene encapsulation. This new 2D material is synthesized directly between graphene layers in a single-step wet-chemical process and is henceforth called hexagonal copper iodide (2D h-CuI). We fully characterize its atomic configuration experimentally and confirm the stability of the obtained heterostructure via density functional theory calculations. The experimental identification of the material is obtained through a combination of scanning TEM (STEM) atomic resolution $Z$-contrast and ptychographic imaging, electron diffraction, X-ray absorption spectroscopy (XAS), spatially resolved electron energy loss spectroscopy (EELS), as well as newly developed few-tilt tomography. We believe that our results demonstrate a route to experimentally access further exotic 2D materials at room temperature.

\subsection{Atomic structure}

\begin{figure}[ht!]
\centering
\includegraphics[width=1\textwidth,keepaspectratio]{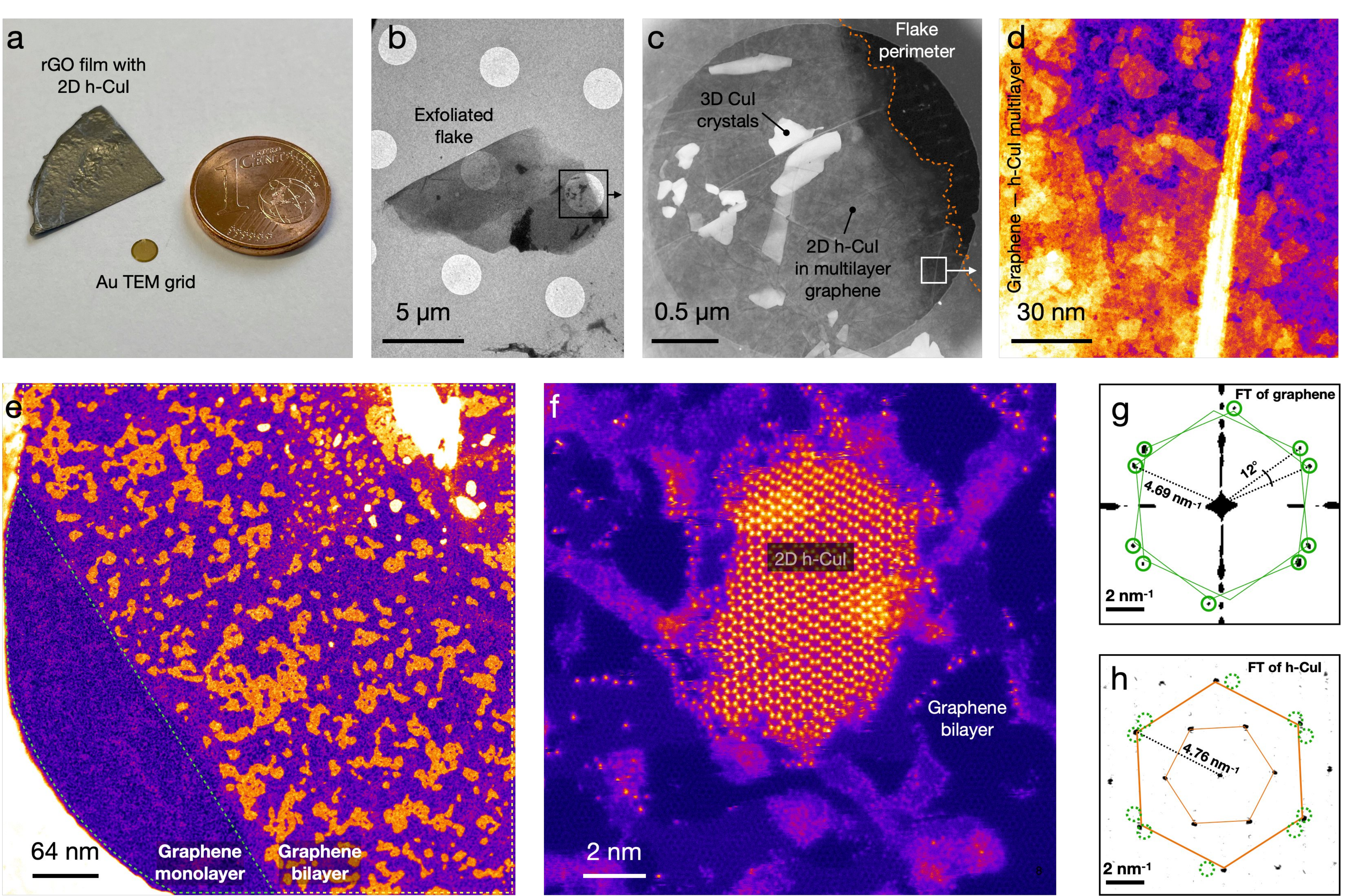}
\caption{(a) A film of reduced graphene oxide (rGO) incorporating 2D h-CuI. (b) Brightfield TEM image of a single graphene/h-CuI flake suspended on a TEM support film. (c-d) High-angle annular dark-field (HAADF) images of the flake edge. The orange and yellow contrast values in d are h-CuI. (e) HAADF overview image of monolayer h-CuI crystals encapsulated in a bilayer graphene sandwich. No h-CuI contrast is visible on the monolayer area on the left-hand side. (f) Atomically-resolved HAADF closeup of a single 2D h-CuI crystal with a magnifying inset in the top right corner. The frequency domain representation of graphene outside the crystal is shown in (g) and that of the area demarcated with a broken line on the crystal in panel (f) is shown in (h). The solid circles in (g) indicate the visible 1st-order graphene reflections, superimposed on (h) using dashed circles. Panels d-f were false-coloured by applying the ImageJ lookup table Fire as an aid to the eye.}
\label{fig:structure}
\end{figure}

Macroscopic quantities of reduced graphene oxide (rGO) encapsulated 2D h-CuI crystals were produced in film form as shown in Figure~\ref{fig:structure}a. The single-step process is described in the Methods section. Note that due to the high quality of the rGO, the individual layers are practically indistinguishable from pure graphene. The layers that were manually exfoliated from the filtrated rGO/h-CuI film and placed on a TEM grid for the STEM/TEM analysis had an average size of ca.~10-20 $\mu$m and a thickness that varied typically between 1-10 layers. An example flake transferred onto a TEM support is displayed in panel b. The flake edge area imaged in STEM high-angle annular dark field (HAADF) mode (panels c and d) show the layered nature of the structure that becomes progressively thinner towards its edge. 
Particularly, the crystallites that are inside the graphene bilayer in panel e, display a uniform contrast over the entire field-of-view, implying they are of uniform thickness. Note that this image was selected because it shows both the monolayer and bilayer graphene areas of an exfoliated flake. Since the crystal islands are randomly dispersed only on the right-hand side of the image on the area covered by the graphene bilayer and the monolayer part on the left is completely devoid of them, it is clear that the crystals are in graphene encapsulation. Additional example images, showing also larger h-CuI domains and higher coverage, can be found in Supplementary Figure~\ref{fig:domains}. The lateral size of the h-CuI single-crystal grains observed in the STEM/TEM images is up to 60 nm; nonetheless, the h-CuI grains are often forming a covalently interconnected network which is extended to areas with a lateral size up to a micrometer. A closer inspection shown in Figure~\ref{fig:structure}f reveals the crystals hexagonally symmetric lattice that matches to the expected symmetry of~$\beta$-CuI.

The periodicity of a lattice in the in-plane direction can be analyzed by converting the projected real-space image into the frequency domain by via a Fourier transform, or alternatively by electron diffraction probing the reciprocal space directly. By selecting the particular areas of interest in Figure~\ref{fig:structure}f, separate Fourier transforms were calculated for the graphene and the 2D h-CuI, as displayed in panels g and h. The period of the second-order h-CuI reflections matches closely the first-order reflections of graphene, implying the commensurability of the lattices; for comparison, the graphene first-order reflections (green circles) are indicated in both panels. The average value of the in-plane lattice parameter of the 2D h-CuI determined from electron diffraction is 4.19$\pm$0.07~\AA, which is slightly smaller than the 4.26~\AA~hexagon-to-hexagon distance along the armchair direction in graphene. Due to the possibility of a minor sample tilt adding a systematic error to the determined absolute values, the spacings were analyzed by directly comparing the graphene reflections with those of the 2D h-CuI within individual nano-beam diffraction images. It is also noteworthy that a slight anisotropy of 1-2\% is observed with the nearest-neighbor distances: in one direction the spacing is usually closely commensurate with the periodicity of graphene, whereas the distances in the perpendicular direction is slightly smaller. These minute differences are indicated in a nano-beam electron diffraction pattern shown in Supplementary Figure~\ref{fig:diffraction}.

The orientations of the graphene layers and the 2D h-CuI crystals were also found to be highly correlated. In all instances, the orientations of the h-CuI crystals match closely with one of the encapsulating graphene lattice directions with a 30\textdegree~rotational translation, although small deviations from the ideal alignment are observable. For instance, the area highlighted with a rectangle in Figure~\ref{fig:structure}f contains (besides h-CuI) bilayer graphene with a layer mismatch angle of 12$\pm$1\textdegree, as is indicated in the Fourier transform in Figure~\ref{fig:structure}g. The graphene armchair edges in this particular case deviate by ca. 6\textdegree~from the horizontal direction of the real-space image, whereas the zigzag direction of the h-CuI crystal deviates by ca. 2.5\textdegree. The edge of a cleaner 2D h-CuI crystal is magnified in Supplementary Figure~\ref{fig:moire}, showing also the crystal orientation with respect to graphene bilayer moiré. Note that the favored alignment is independent of the CuI crystallite size, but small crystals are sometimes seen oscillating between the two graphene orientations during image acquisition, indicating relatively weak inter-layer binding. Even rotation angles of up to 30{\textdegree} are possible between two consecutive images. Examples are shown in Supplementary Figure~\ref{fig:rotating} and Video S1.

\subsection{Elemental composition}

\begin{figure}[ht!]
\centering
\includegraphics[width=1\textwidth,keepaspectratio]{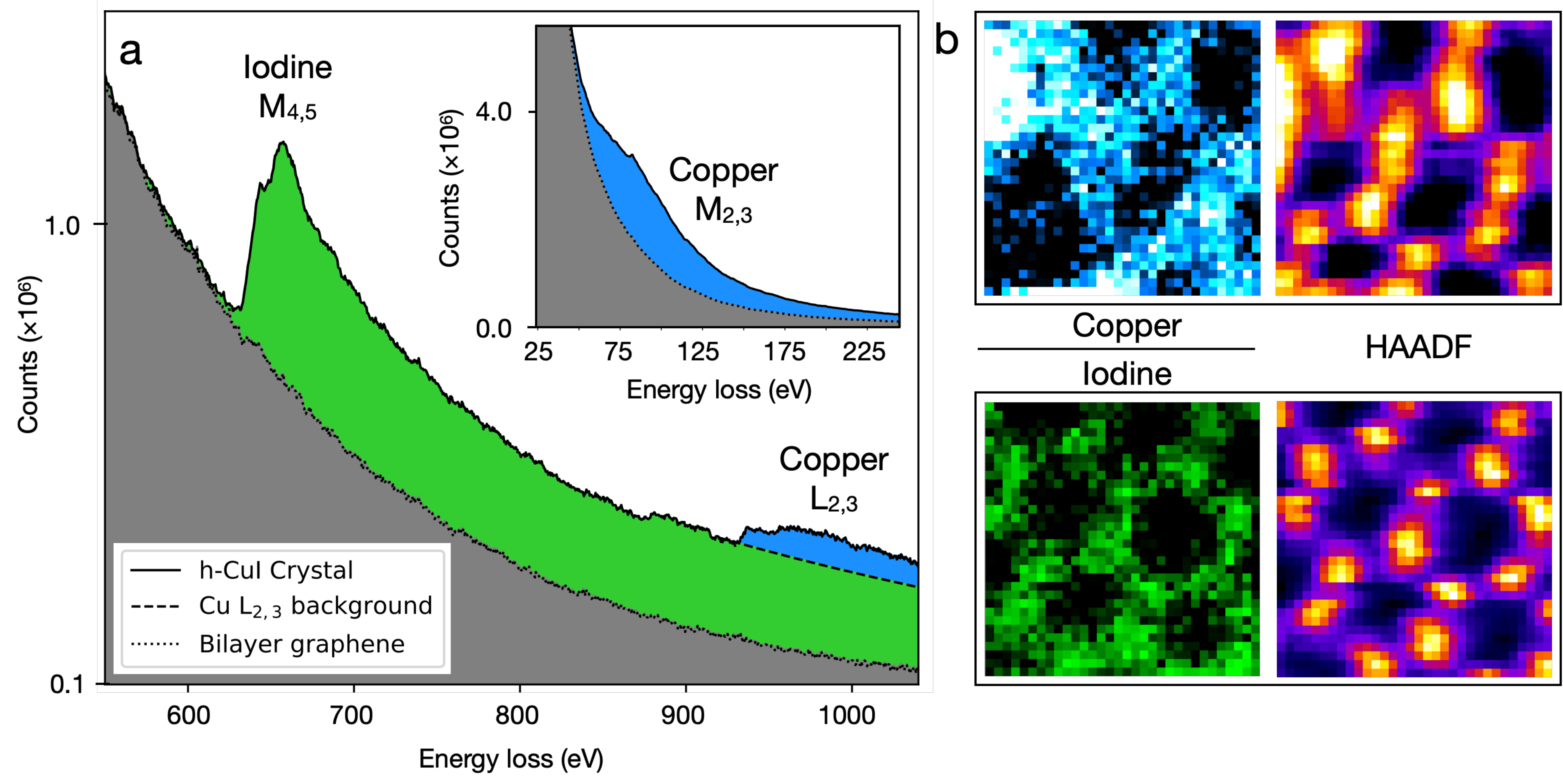}
\caption{(a) Iodine and copper electron energy core-loss spectra recorded from a 2D h-CuI crystal similar to the one shown in Figure~\ref{fig:structure}f. The bilayer graphene reference was acquired from outside the crystal. The Cu~$M_{2,3}$ core-loss edge is shown in the inset. (b) Spatially resolved EELS maps showing Cu (integrated over 65-185 eV) and I (630-750 eV) distributions in the crystal. The distortion in the images is caused by the sample-stage drift during the ca.~3 minutes of data acquisition required to achieve the required signal-to-noise-ratio. The greater variation on the Cu EELS map signal intensities is likely a result of lower electron-beam stability of the Cu atomic sites. The field-of-view of the maps is 1$\times$1\,nm$^2$. The right-most panels show the concurrently acquired HAADF signals in false-colour (ImageJ lookup table Fire).}
\label{fig:eels}
\end{figure}

\begin{figure}[ht!]
\centering
\includegraphics[width=0.65\textwidth,keepaspectratio]{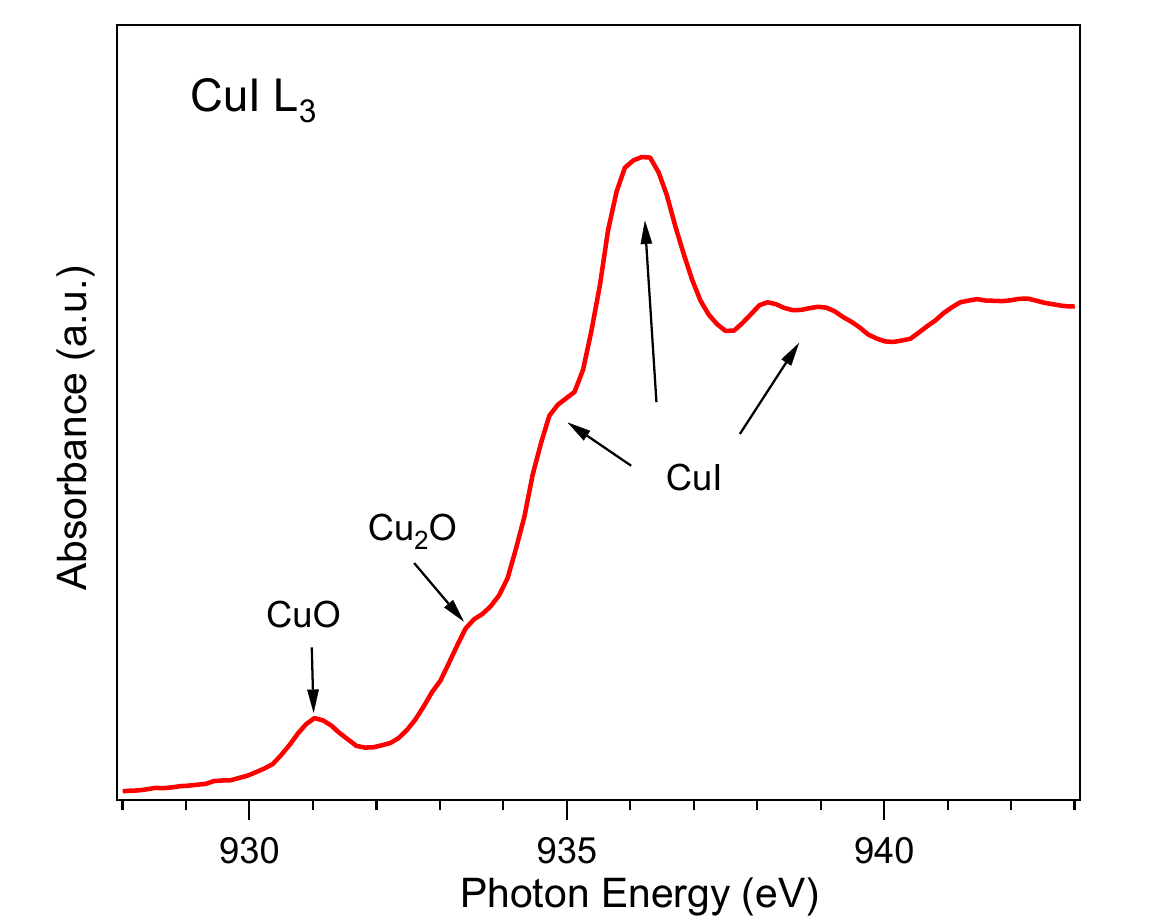}
\caption{X-ray absorption spectra of CuI near the Cu $L_3$-edge. The arrows indicate different spectral features as explained in the text.}
\label{fig:xas}
\end{figure}

The elemental composition of the crystals was analyzed via EELS and the oxidation state of the elements via XAS. The results are summarized in Figures~\ref{fig:eels} and~\ref{fig:xas}. Figure~\ref{fig:eels}a shows an energy-loss spectrum acquired from the area of 4 x 4 nm$^2$ while  continuously scanning the electron probe over a single h-CuI crystal. The two core-loss edges visible at 619\,eV and 931\,eV are  associated with the I-$M$ and Cu-$L$ shell electron excitation. The pristine bilayer graphene spectrum (gray area) was recorded next to the h-CuI crystal as a reference. The inset of Figure~\ref{fig:eels}a shows the $M_{2,3}$ core-loss edge of copper. The spatial distributions of the core-loss signal sources are presented as elemental maps in Figure~\ref{fig:eels}b, and are well commensurate with the simultaneously acquired ADF-signal. This result demonstrates that each of the apparent atomic positions is, in fact, occupied by at least a single I and Cu atom that are stacked exactly on top of one another as would be expected in $\beta$-CuI.\cite{sakuma1988crystal, keen1994determination} 

The Cu $L_3$ XAS spectrum shown in Figure~\ref{fig:xas} allows us to clearly identify the CuI compound thanks to several characteristic spectral features indicated by arrows, reproducing well the results from literature \cite{hayakawa2021x, meulenberg2004structure}. The main absorption maximum at around 936 eV is assigned to the 2p $\rightarrow$ 4s (2p$^6$3d$^{10}$  $\rightarrow$ 2p$^5$3d$^{10}$4s$^1$) transition combined with 3d-4s hybridization as detailed in Ref. \citenum{hayakawa2021x}. The shoulder close to 935 eV is part of the CuI characteristic line-shape and has been attributed to a valence exciton \cite{meulenberg2004structure}.

Two other features are present close to photon energies of 931 and 933 eV that we attribute to CuO and Cu$_2$O compounds, respectively. Both copper oxides probably originate from the carbon tape, which contains them as we measured separately (the results are not shown here). As our CuI sample is a rather porous film, the signal from the supporting carbon tape can presumably add to the overall measured intensity. The white line of the iodine $M_2$-edge is certainly present as well but superimposed by CuO.

\subsection{Atomic structure in cross section}

\begin{figure}[ht!]
\centering
\includegraphics[width=1.0\textwidth,keepaspectratio]{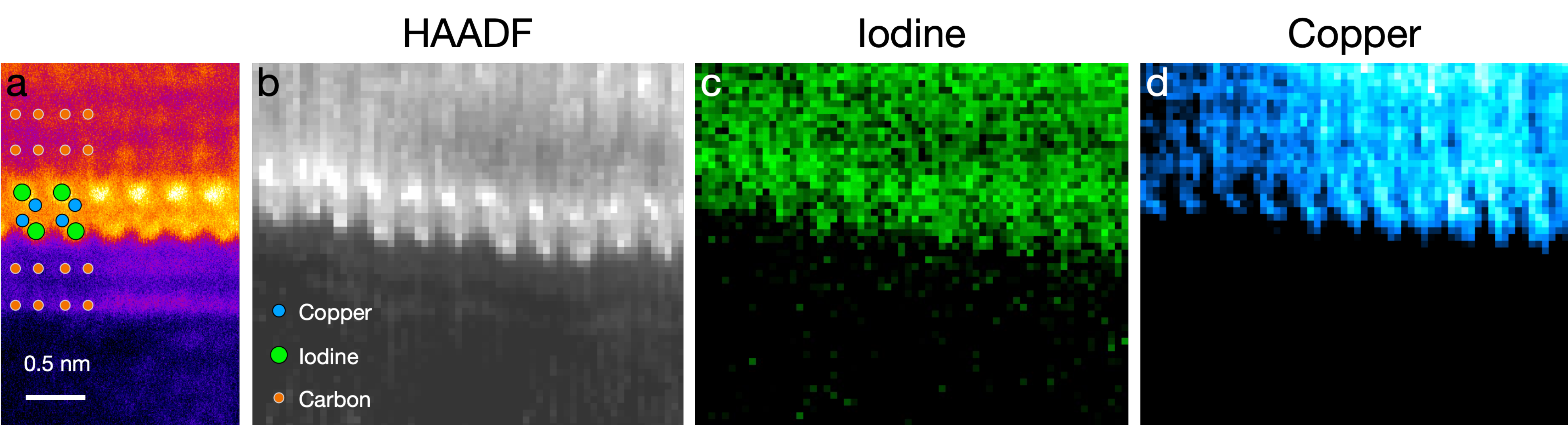}
\caption{Cross-sectional STEM images and EELS maps of the 2D h-CuI structure within multilayer graphene encapsulation. The atomic layers are indicated by the labels in panels (a-b), of which the first one is an average of four rapid-scanned STEM HAADF frames to improve the contrast of graphene. (c-d) show the iodine and copper $M$-edge intensities and panel (b) the simultaneously acquired HAADF signal. The iodine sites on the top row of the HAADF panels appear slightly brighter likely due to a greater overlap of the atoms on those atomic columns. The elemental EELS maps on the right were produced by integrating the Cu-$M$ edge (65-185 eV) and I-$M$ edge (630-750 eV) signal intensities. The left panel is shown in false-colour (ImageJ lookup table Fire).}
\label{fig:Cross-section}
\end{figure}

Three complementary techniques were further applied to study the 3D atomic structure: cross-sectional imaging~\cite{haigh2012cross}, observations of shallow-angle tilted projections, including electron diffraction, and a recently developed method capable of reconstructing arbitrary 2D materials based on few-tilt tomography~\cite{hofer2021threedimensional}. Cross-sectional images provide the most direct proof of the atomic configuration and are presented in Figure~\ref{fig:Cross-section}. Here, a folded graphene/h-CuI heterostructure that includes several graphene layers crossing the image focal plane in profile was observed.

Figure~\ref{fig:Cross-section}a displays a high-resolution HAADF image of a well-resolved CuI layer between less visible layers of graphene: the top and bottom rows of bright atoms in the central 2D structure are likely iodine, matching the iodine-terminating configuration for a monolayer of $\beta$-CuI~\cite{sakuma1988crystal, mounet2018two}. A schematic model of $\beta$-CuI is also draw on top of the structure. To visualize the distribution of iodine and copper atoms in the cross-section, EELS maps were acquired by integrating iodine I-$M$ (630-750 eV) and copper Cu-$M$ (65-185 eV) edge intensities that are displayed in Figure~\ref{fig:Cross-section}c-d. A simultaneously acquired STEM HAADF image is shown in Figure~\ref{fig:Cross-section}b. Note that the the EELS signals above the crystals edge originates from the out-of-focus part of the bent flake, which is schematically depicted in Figure~\ref{fig:bent_flake}. The fine-structure of the crystal and even the I positions are still discernible in the middle part of panel c. The distortion in the EELS maps, which is also visible in the simultaneously acquired HAADF image, results mainly from the sample stage drift during ca. 10 minutes of data acquisition, with a possible contribution from structural dynamics induced by the electron irradiation.

Additional evidence of the monolayer nature is obtained through observations of the structure from different viewing angles, with a single h-CuI crystal being rotated around a pair of perpendicularly aligned tilt axes at $\pm$17\textdegree~angles. The resulting images are shown in Figure~\ref{fig:STEM tilt}, and a larger field of view of the area can be found in Supplementary Figure~\ref{fig:domains} and images recorded for tilts along the second tilt axis are shown in Figure~\ref{fig:tilt_SI}. To interpret the experimental images, we simulated a number of views (see Methods) with the same tilt angles as in the experiment and compared them with the experimental data. The simulated images are based on the optimized h-CuI monolayer structure acquired via density functional theory (DFT) calculations discussed in more detail below. The bilayer image simulations were based on a DFT energy optimized $\beta$-CuI structure with an out-of-plane unit cell length of 7.345~\AA~(see Table S1). Whereas the simulated bilayer matches poorly with the experimental observations, the monolayer model is an excellent match in all projections. This result was further corroborated by electron diffraction tilt experiments that are discussed in the Supplementary Material (see especially Supplementary Figure~\ref{fig:DPtomo}).

\begin{figure}[ht!]
\centering
\includegraphics[width=1.0\textwidth,keepaspectratio]{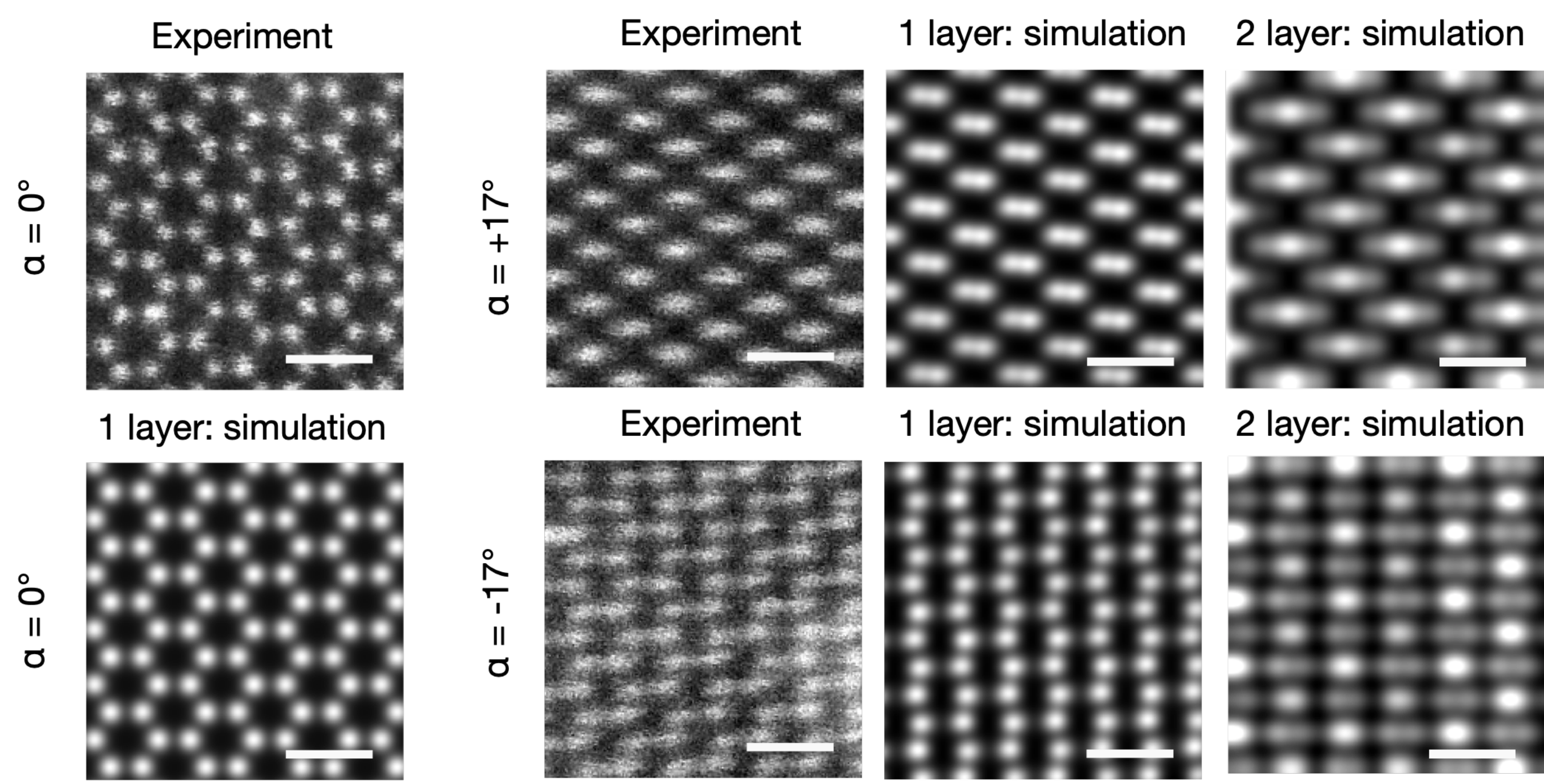}
\caption{STEM-HAADF projections of a h-CuI crystal from different viewing angles shown with simulated views of a monolayer and AA-stacked bilayer structures. The images on the left panels show the case of normal electron beam incidence where only the monolayer structure is considered (AA-bilayer would display identical symmetry in the normal projection). The following frames compare the experimental and simulated projections along a single tilt axis $\alpha$ = $\pm$17\textdegree. The image simulations are based on the density functional theory models shown in Figure~\ref{fig:models} and in Figure~\ref{fig:bulk_cui}. The bilayer out-of-plane unit cell length in the simulations was 7.345~\AA. The scale bars in the images are 6.5~\AA.}
\label{fig:STEM tilt}
\end{figure}

With all evidence thus far pointing to a monolayer structure we now turn to the extraction of the 3D coordinates of Cu and I atoms within an individual h-CuI layer. Few tilt tomography has previously enabled the 3D reconstruction of graphene with ADF STEM images from as few as two tilt angles~\cite{hofer2018revealing}. The CuI system however is far more challenging because it is not only several atomic layers thick, but also contains a mixture of heavy and light elements in each atomic column. Light elements are typically obscured in ADF images by the neighboring heavy atoms, as they indeed also do in the tilted projections here, and thus we complement the ADF signal with simultaneous single side band (SSB) ptychography~\cite{Pennycook2015}. Ptychography provides a greatly enhanced signal for a given electron dose compared to ADF imaging~\cite{PENNYCOOK2019131}, and simultaneously resolves both heavy and light elements~\cite{Yang2016}. Details of this analysis are given in Methods and in the Supplementary Material. A more complete description of the method can be found in Ref.~\citenum{hofer2021threedimensional}.    

\begin{figure}[ht!]
\centering
\includegraphics[width=1.0\textwidth,keepaspectratio]{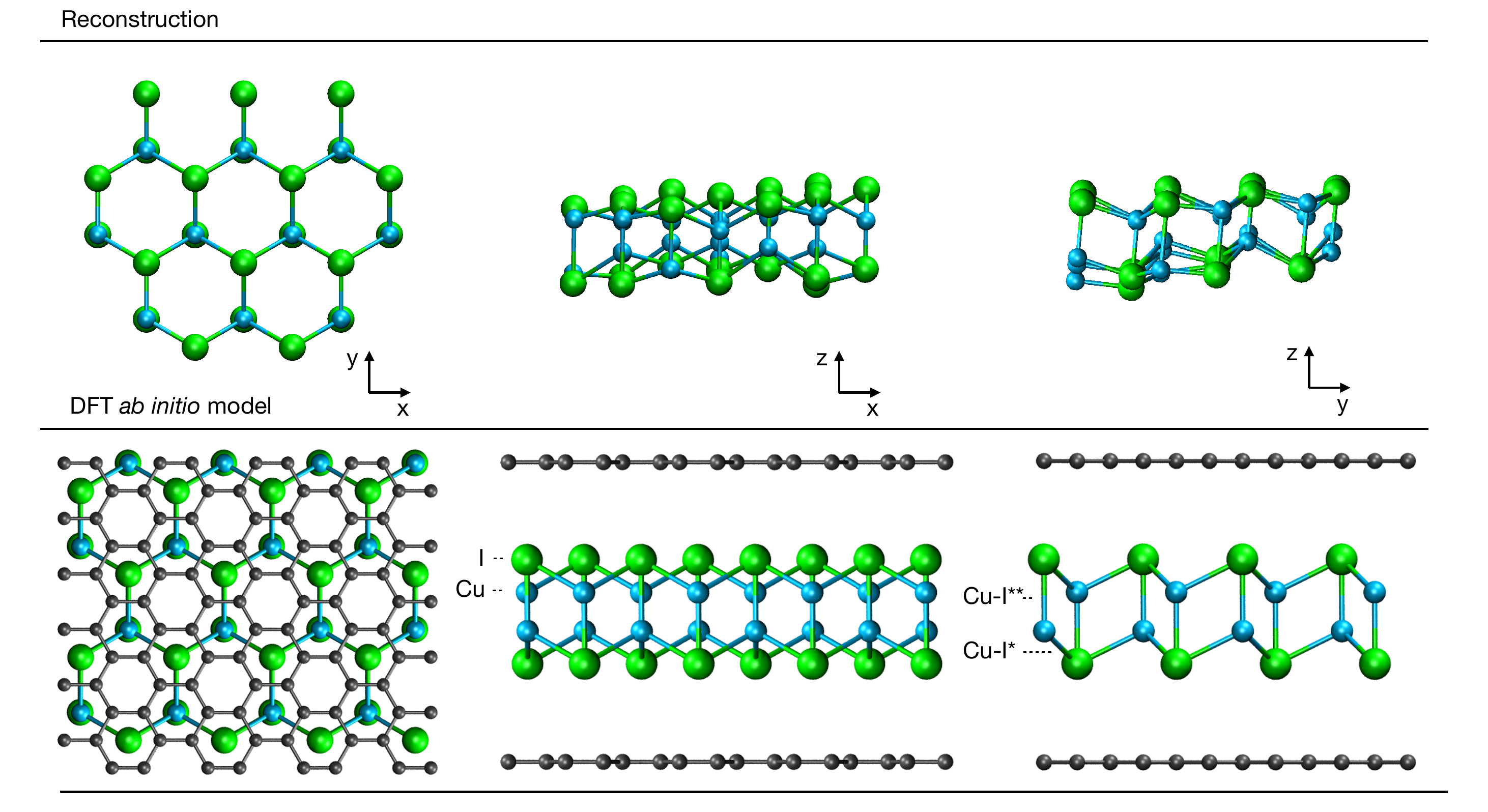}
\caption{Reconstructed 2D h-CuI crystal based on tilt experiments on the top row and an \textit{ab initio} optimized structure in bilayer graphene encapsulation below it. The projections which the reconstruction is based on are shown in Supplementary Figure~\ref{fig:SSB}. The graphene layers in the DFT model are aligned in AA-stacking. The DFT calculations are based on the optPBE-vdW functional~\cite{klime_2009}.}
\label{fig:models}
\end{figure}

Figure~\ref{fig:models} shows a ball-and-stick model of 2D h-CuI reconstructed from the experimental few tilt tomography together with an optimized structure obtained from the DFT calculations (see Methods). Due to the convolution of the atomic positions with the atomic vibrations perpendicular to the plane, some uncertainty in the tomographic result is expected, especially in the out-of-plane direction.

With some minute differences, however, the experimental and computational models are in good agreement. For instance, the $z$-displacement between the terminating iodine atomic planes in the \textit{ab initio} model was 3.89~{\AA}, whereas a separation of 3.42$\pm$0.34~{\AA}~(mean and one standard deviation) was extracted from the reconstruction. The experimentally determined 1.63$\pm$0.35~{\AA} distance between the copper atomic planes, meanwhile, is slightly larger than the DFT result of 1.45~{\AA}. The experimental Cu--I bond lengths were 2.67$\pm$0.16~{\AA} in the in-plane direction and 2.55$\pm$0.49~{\AA} in the out-of-plane direction, whereas the DFT equivalents were 2.75~{\AA} and 2.67~{\AA}. These particular atomic bonds are also indicated in Figure~\ref{fig:models} as Cu--I$^{*}$ and Cu--I$^{**}$, respectively.

The DFT-optimized separation of the graphene and CuI layers is 3.678~{\AA} with a very small variation of~$\pm$0.013~{\AA} depending on the interlayer stacking configuration (see Table S3). In the most energetically favourable DFT configuration, iodine atoms are facing 
the centers of the graphene hexagons. The binding energies were calculated to be 14.1$\pm$0.1~meV/{\AA}$^2$ per graphene layer, which is slightly higher than the calculated interlayer binding energy in bulk $\beta$-CuI of 13.4~meV/{\AA}$^2$. A single layer of $\beta$-CuI has been predicted to be a direct band gap semiconductor~\cite{mounet2018two, huang2020group}. Our calculations with the HSE06 hybrid functional give a band gap of 3.17~eV for a single layer and 2.68~eV for bulk $\beta$-CuI. Band structure calculations further suggest that in the CuI/graphene heterostructure, the characteristic linear dispersion of graphene $\pi$-bands is preserved and the Dirac point appears in the band gap of CuI (see Supplementary Figure~\ref{fig:bs_gr_cui} and Table~S2).

\subsection{Discussion}

The structures of different cuprous iodide phases have in the past been subject to some level of controversy. Excluding the 2D phase presented here and the 1D structures observed in carbon nanotubes~\cite{kiselev2013structure}, a total of eight solid phases have thus far been identified~\cite{hull1998superionic, keen1994determination}. The superionic phase occurring at high-temperatures particularly follows from the chemical bonding of copper and iodine that is neither purely of covalent nor ionic nature, but exhibits both characteristics simultaneously~\cite{phillips1970ionicity}. Normally, the semiconducting zinc blende bulk crystal~\cite{ahn2016cuprous} is the primary phase in ambient conditions and it transforms into the layered $\beta$-CuI phase at ca.~645-647~K~\cite{sakuma1988crystal, keen1994determination, hull1998superionic}, and which we presently obtained also at room temperature in 2D form by graphene encapsulation. The aforementioned controversy emerged in 1952 when the vdW-layered $\beta$-CuI phase was simultaneously but erroneously identified as a wurtzite structure by two independent research groups~\cite{miyake1952phase, KrugSieg1952}. This error was pointed out by Bührer and Hälg in 1977~\cite{buhrer1977crystal}, but since the authors did not provide the correct unit cell description, the confusion remained and the wurtzite structure sometimes resurfaces in the scientific literature~\cite{liu2019ideal}. However, first principle calculations imply that the wurtzite phase might, after all, exist near the ground state~\cite{hernandez2009density}. To our knowledge, the $\beta$-CuI phase was first described by Sakuma in 1988 and only later confirmed by Keen and Hull~\cite{keen1994determination}, and Sakuma should be rightly credited for this contribution~\cite{sakuma1988crystal}.

The size of the 2D single crystal domains we observe span from a few nanometers to some tens of nanometers and exhibit high stability under 60~keV electron-beam exposure in ultra-high vacuum conditions. Although interconnected islands often cover areas as large as micrometers across, significant intrinsic disorder in the form of grain boundaries and nano-pores is typically observed in them (see e.g.~Supplementary Figure~\ref{fig:domains}). We believe that the crystal size might be limited by the slight incommensurability of the graphene and h-CuI lattices, reflected in the measured anisotropy for the different lattice directions. This likely results in a small but significant strain energy penalty that can be sustained only for a finite crystal domain size, as well as through the appearance of nano-pores and voids. The in-plane lattice constant of the 2D h-CuI crystal measured here is also somewhat smaller than what was previously reported for the bulk crystals. Sakuma measured a value of 4.279~{\AA} for the $\beta$-CuI phase at 693~K using x-rays and Keen and Hull 4.304~{\AA} at 655~K in neutron diffraction experiments, whereas we measured 4.19$\pm$0.07~{\AA} for a monolayer by using nano-beam electron diffraction at room temperature. This apparent discrepancy might be due to thermal expansion at temperatures normally required to stabilize the $\beta$-CuI phase (a spacing of ca. 4.21~{\AA} at room temperature can be extrapolated based on the thermal expansion coefficient defined by Keen and Hull). This value also matches our DFT calculations that predict a 4.20~{\AA} spacing for both a monolayer and a bulk crystal without the graphene envelope (see Supplementary Figure~\ref{fig:bulk_cui} and Table S1). As was also pointed out by Sakuma, each atom in this particular crystal structure has exactly three neighbors and the crystal belongs to the p$\overline{3}$m1 Hermann Mauguin space group. To our knowledge, the h-CuI crystal is one of few 1:1 stoichiometric 2D materials within this particular symmetry (hexagonal boron nitride being the most well known), and, besides PbI$_2$~\cite{sinha2020atomic}, the only one that incorporates also a halogen atom.

The discovery of 2D h-CuI is particularly interesting, since magnons~\cite{klein2018probing} and layer\hyp{}structure\hyp{}dependent ferromagnetism have recently been reported in CrI$_3$~\cite{huang2017layer, kashin2020orbitally, wang2021systematic}, and a host of further magnetic ordering phenomena have been predicted in other 2D metal\hyp{}halides such as in NiI$_2$ and CoI$_2$~\cite{mounet2018two, huang2020group, amoroso2020spontaneous, kurumaji2013magnetoelectric}. Although DFT simulations have suggested the possibility of exfoliating some of those compounds, including 2D h-CuI~\cite{mounet2018two}, their thermodynamic stability under ambient conditions remains untested. Indeed, based on the experiments by Sakuma~\cite{sakuma1988crystal} and Keen and Hull~\cite{keen1994determination}, it seems obvious that 2D h-CuI would not be stable at room temperature without encapsulation. The stabilizing function of graphene encapsulation was also manifested in our high-resolution TEM experiments: after 80~keV electrons breached one of the graphene layers, the 2D h-CuI rapidly broke apart (see Supplementary Figures~\ref{fig:rotating} and~\ref{fig:cuo}, and Video S1).

Finally, there is no reason to believe that the present synthesis approach, which was here used to stabilize crystals of 2D cuprous iodide for room temperature experiments, would be limited to this particular structure. Quite the contrary, we have already applied the same process to produce 2D silver iodide (AgI) and nickel iodide (NiI$_2$) crystals, both of which are shown in Figure~\ref{fig:metal_halides}, and generalizing this concept should allow access to further exotic layered structures vastly expanding the currently available library of 2D materials, and their incorporation in devices and their structure and properties to be studied in room temperature.

\newpage

\bibliography{2DCuI}

\providecommand{\latin}[1]{#1}
\makeatletter
\providecommand{\doi}
  {\begingroup\let\do\@makeother\dospecials
  \catcode`\{=1 \catcode`\}=2 \doi@aux}
\providecommand{\doi@aux}[1]{\endgroup\texttt{#1}}
\makeatother
\providecommand*\mcitethebibliography{\thebibliography}
\csname @ifundefined\endcsname{endmcitethebibliography}
  {\let\endmcitethebibliography\endthebibliography}{}
\begin{mcitethebibliography}{54}
\providecommand*\natexlab[1]{#1}
\providecommand*\mciteSetBstSublistMode[1]{}
\providecommand*\mciteSetBstMaxWidthForm[2]{}
\providecommand*\mciteBstWouldAddEndPuncttrue
  {\def\EndOfBibitem{\unskip.}}
\providecommand*\mciteBstWouldAddEndPunctfalse
  {\let\EndOfBibitem\relax}
\providecommand*\mciteSetBstMidEndSepPunct[3]{}
\providecommand*\mciteSetBstSublistLabelBeginEnd[3]{}
\providecommand*\EndOfBibitem{}
\mciteSetBstSublistMode{f}
\mciteSetBstMaxWidthForm{subitem}{(\alph{mcitesubitemcount})}
\mciteSetBstSublistLabelBeginEnd
  {\mcitemaxwidthsubitemform\space}
  {\relax}
  {\relax}

\bibitem[Novoselov \latin{et~al.}(2004)Novoselov, Geim, Morozov, Jiang, Zhang,
  Dubonos, Grigorieva, and Firsov]{novoselov2004electric}
Novoselov,~K.~S.; Geim,~A.~K.; Morozov,~S.~V.; Jiang,~D.; Zhang,~Y.;
  Dubonos,~S.~V.; Grigorieva,~I.~V.; Firsov,~A.~A. Electric field effect in
  atomically thin carbon films. \emph{Science} \textbf{2004}, \emph{306},
  666--669\relax
\mciteBstWouldAddEndPuncttrue
\mciteSetBstMidEndSepPunct{\mcitedefaultmidpunct}
{\mcitedefaultendpunct}{\mcitedefaultseppunct}\relax
\EndOfBibitem
\bibitem[Huang \latin{et~al.}(2012)Huang, Kurasch, Srivastava, Skakalova,
  Kotakoski, Krasheninnikov, Hovden, Mao, Meyer, Smet, \latin{et~al.}
  others]{huang2012direct}
Huang,~P.~Y.; Kurasch,~S.; Srivastava,~A.; Skakalova,~V.; Kotakoski,~J.;
  Krasheninnikov,~A.~V.; Hovden,~R.; Mao,~Q.; Meyer,~J.~C.; Smet,~J.,
  \latin{et~al.}  Direct imaging of a two-dimensional silica glass on graphene.
  \emph{Nano Letters} \textbf{2012}, \emph{12}, 1081--1086\relax
\mciteBstWouldAddEndPuncttrue
\mciteSetBstMidEndSepPunct{\mcitedefaultmidpunct}
{\mcitedefaultendpunct}{\mcitedefaultseppunct}\relax
\EndOfBibitem
\bibitem[Sinha \latin{et~al.}(2020)Sinha, Zhu, France-Lanord, Sheng, Grossman,
  Porfyrakis, and Warner]{sinha2020atomic}
Sinha,~S.; Zhu,~T.; France-Lanord,~A.; Sheng,~Y.; Grossman,~J.~C.;
  Porfyrakis,~K.; Warner,~J.~H. Atomic structure and defect dynamics of
  monolayer lead iodide nanodisks with epitaxial alignment on graphene.
  \emph{Nature Communications} \textbf{2020}, \emph{11}, 1--13\relax
\mciteBstWouldAddEndPuncttrue
\mciteSetBstMidEndSepPunct{\mcitedefaultmidpunct}
{\mcitedefaultendpunct}{\mcitedefaultseppunct}\relax
\EndOfBibitem
\bibitem[Zagler \latin{et~al.}(2020)Zagler, Reticcioli, Mangler, Scheinecker,
  Franchini, and Kotakoski]{zagler2020cuau}
Zagler,~G.; Reticcioli,~M.; Mangler,~C.; Scheinecker,~D.; Franchini,~C.;
  Kotakoski,~J. CuAu, a hexagonal two-dimensional metal. \emph{2D Materials}
  \textbf{2020}, \emph{7}, 045017\relax
\mciteBstWouldAddEndPuncttrue
\mciteSetBstMidEndSepPunct{\mcitedefaultmidpunct}
{\mcitedefaultendpunct}{\mcitedefaultseppunct}\relax
\EndOfBibitem
\bibitem[Al~Balushi \latin{et~al.}(2016)Al~Balushi, Wang, Ghosh, Vil{\'a},
  Eichfeld, Caldwell, Qin, Lin, DeSario, Stone, \latin{et~al.}
  others]{al2016two}
Al~Balushi,~Z.~Y.; Wang,~K.; Ghosh,~R.~K.; Vil{\'a},~R.~A.; Eichfeld,~S.~M.;
  Caldwell,~J.~D.; Qin,~X.; Lin,~Y.-C.; DeSario,~P.~A.; Stone,~G.,
  \latin{et~al.}  Two-dimensional gallium nitride realized via graphene
  encapsulation. \emph{Nature materials} \textbf{2016}, \emph{15},
  1166--1171\relax
\mciteBstWouldAddEndPuncttrue
\mciteSetBstMidEndSepPunct{\mcitedefaultmidpunct}
{\mcitedefaultendpunct}{\mcitedefaultseppunct}\relax
\EndOfBibitem
\bibitem[P{\'e}cz \latin{et~al.}(2021)P{\'e}cz, Nicotra, Giannazzo, Yakimova,
  Koos, and Kakanakova-Georgieva]{pecz2021indium}
P{\'e}cz,~B.; Nicotra,~G.; Giannazzo,~F.; Yakimova,~R.; Koos,~A.;
  Kakanakova-Georgieva,~A. Indium Nitride at the 2D Limit. \emph{Advanced
  Materials} \textbf{2021}, \emph{33}, 2006660\relax
\mciteBstWouldAddEndPuncttrue
\mciteSetBstMidEndSepPunct{\mcitedefaultmidpunct}
{\mcitedefaultendpunct}{\mcitedefaultseppunct}\relax
\EndOfBibitem
\bibitem[Rosenzweig and Starke(2020)Rosenzweig, and
  Starke]{rosenzweig2020large}
Rosenzweig,~P.; Starke,~U. Large-area synthesis of a semiconducting silver
  monolayer via intercalation of epitaxial graphene. \emph{Physical Review B}
  \textbf{2020}, \emph{101}, 201407\relax
\mciteBstWouldAddEndPuncttrue
\mciteSetBstMidEndSepPunct{\mcitedefaultmidpunct}
{\mcitedefaultendpunct}{\mcitedefaultseppunct}\relax
\EndOfBibitem
\bibitem[Forti \latin{et~al.}(2020)Forti, Link, St{\"o}hr, Niu, Zakharov,
  Coletti, and Starke]{forti2020semiconductor}
Forti,~S.; Link,~S.; St{\"o}hr,~A.; Niu,~Y.; Zakharov,~A.~A.; Coletti,~C.;
  Starke,~U. Semiconductor to metal transition in two-dimensional gold and its
  van der Waals heterostack with graphene. \emph{Nature communications}
  \textbf{2020}, \emph{11}, 1--7\relax
\mciteBstWouldAddEndPuncttrue
\mciteSetBstMidEndSepPunct{\mcitedefaultmidpunct}
{\mcitedefaultendpunct}{\mcitedefaultseppunct}\relax
\EndOfBibitem
\bibitem[Briggs \latin{et~al.}(2020)Briggs, Bersch, Wang, Jiang, Koch, Nayir,
  Wang, Kolmer, Ko, Duran, \latin{et~al.} others]{briggs2020atomically}
Briggs,~N.; Bersch,~B.; Wang,~Y.; Jiang,~J.; Koch,~R.~J.; Nayir,~N.; Wang,~K.;
  Kolmer,~M.; Ko,~W.; Duran,~A. D. L.~F., \latin{et~al.}  Atomically thin
  half-van der Waals metals enabled by confinement heteroepitaxy. \emph{Nature
  materials} \textbf{2020}, \emph{19}, 637--643\relax
\mciteBstWouldAddEndPuncttrue
\mciteSetBstMidEndSepPunct{\mcitedefaultmidpunct}
{\mcitedefaultendpunct}{\mcitedefaultseppunct}\relax
\EndOfBibitem
\bibitem[K{\"u}hne \latin{et~al.}(2018)K{\"u}hne, B{\"o}rrnert, Fecher,
  Ghorbani-Asl, Biskupek, Samuelis, Krasheninnikov, Kaiser, and
  Smet]{kuhne2018reversible}
K{\"u}hne,~M.; B{\"o}rrnert,~F.; Fecher,~S.; Ghorbani-Asl,~M.; Biskupek,~J.;
  Samuelis,~D.; Krasheninnikov,~A.~V.; Kaiser,~U.; Smet,~J.~H. Reversible
  superdense ordering of lithium between two graphene sheets. \emph{Nature}
  \textbf{2018}, \emph{564}, 234--239\relax
\mciteBstWouldAddEndPuncttrue
\mciteSetBstMidEndSepPunct{\mcitedefaultmidpunct}
{\mcitedefaultendpunct}{\mcitedefaultseppunct}\relax
\EndOfBibitem
\bibitem[Kelly \latin{et~al.}(2018)Kelly, Zhou, Clark, Hamer, Lewis, Rakowski,
  Haigh, and Gorbachev]{kelly2018nanometer}
Kelly,~D.~J.; Zhou,~M.; Clark,~N.; Hamer,~M.~J.; Lewis,~E.~A.; Rakowski,~A.~M.;
  Haigh,~S.~J.; Gorbachev,~R.~V. Nanometer resolution elemental mapping in
  graphene-based TEM liquid cells. \emph{Nano Letters} \textbf{2018},
  \emph{18}, 1168--1174\relax
\mciteBstWouldAddEndPuncttrue
\mciteSetBstMidEndSepPunct{\mcitedefaultmidpunct}
{\mcitedefaultendpunct}{\mcitedefaultseppunct}\relax
\EndOfBibitem
\bibitem[Zan \latin{et~al.}(2013)Zan, Ramasse, Jalil, Georgiou, Bangert, and
  Novoselov]{zan2013control}
Zan,~R.; Ramasse,~Q.~M.; Jalil,~R.; Georgiou,~T.; Bangert,~U.; Novoselov,~K.~S.
  Control of radiation damage in MoS2 by graphene encapsulation. \emph{ACS
  Nano} \textbf{2013}, \emph{7}, 10167--10174\relax
\mciteBstWouldAddEndPuncttrue
\mciteSetBstMidEndSepPunct{\mcitedefaultmidpunct}
{\mcitedefaultendpunct}{\mcitedefaultseppunct}\relax
\EndOfBibitem
\bibitem[Algara-Siller \latin{et~al.}(2013)Algara-Siller, Kurasch, Sedighi,
  Lehtinen, and Kaiser]{algara2013pristine}
Algara-Siller,~G.; Kurasch,~S.; Sedighi,~M.; Lehtinen,~O.; Kaiser,~U. The
  pristine atomic structure of MoS2 monolayer protected from electron radiation
  damage by graphene. \emph{Applied Physics Letters} \textbf{2013}, \emph{103},
  203107\relax
\mciteBstWouldAddEndPuncttrue
\mciteSetBstMidEndSepPunct{\mcitedefaultmidpunct}
{\mcitedefaultendpunct}{\mcitedefaultseppunct}\relax
\EndOfBibitem
\bibitem[Mirzayev \latin{et~al.}(2017)Mirzayev, Mustonen, Monazam,
  Mittelberger, Pennycook, Mangler, Susi, Kotakoski, and
  Meyer]{mirzayev2017buckyball}
Mirzayev,~R.; Mustonen,~K.; Monazam,~M.~R.; Mittelberger,~A.; Pennycook,~T.~J.;
  Mangler,~C.; Susi,~T.; Kotakoski,~J.; Meyer,~J.~C. Buckyball sandwiches.
  \emph{Science Advances} \textbf{2017}, \emph{3}, e1700176\relax
\mciteBstWouldAddEndPuncttrue
\mciteSetBstMidEndSepPunct{\mcitedefaultmidpunct}
{\mcitedefaultendpunct}{\mcitedefaultseppunct}\relax
\EndOfBibitem
\bibitem[L{\"a}ngle \latin{et~al.}(2020)L{\"a}ngle, Mizohata, {\AA}hlgren,
  Trentino, Mustonen, and Kotakoski]{langle20202d}
L{\"a}ngle,~M.; Mizohata,~K.; {\AA}hlgren,~E.~H.; Trentino,~A.; Mustonen,~K.;
  Kotakoski,~J. 2D Noble Gas Crystals Encapsulated in Few-layer Graphene.
  \emph{Microscopy and Microanalysis} \textbf{2020}, \emph{26},
  1086--1089\relax
\mciteBstWouldAddEndPuncttrue
\mciteSetBstMidEndSepPunct{\mcitedefaultmidpunct}
{\mcitedefaultendpunct}{\mcitedefaultseppunct}\relax
\EndOfBibitem
\bibitem[Vasu \latin{et~al.}(2016)Vasu, Prestat, Abraham, Dix, Kashtiban,
  Beheshtian, Sloan, Carbone, Neek-Amal, Haigh, \latin{et~al.}
  others]{vasu2016van}
Vasu,~K.; Prestat,~E.; Abraham,~J.; Dix,~J.; Kashtiban,~R.; Beheshtian,~J.;
  Sloan,~J.; Carbone,~P.; Neek-Amal,~M.; Haigh,~S., \latin{et~al.}  Van der
  Waals pressure and its effect on trapped interlayer molecules. \emph{Nature
  Communications} \textbf{2016}, \emph{7}, 1--6\relax
\mciteBstWouldAddEndPuncttrue
\mciteSetBstMidEndSepPunct{\mcitedefaultmidpunct}
{\mcitedefaultendpunct}{\mcitedefaultseppunct}\relax
\EndOfBibitem
\bibitem[Mounet \latin{et~al.}(2018)Mounet, Gibertini, Schwaller, Campi,
  Merkys, Marrazzo, Sohier, Castelli, Cepellotti, Pizzi, \latin{et~al.}
  others]{mounet2018two}
Mounet,~N.; Gibertini,~M.; Schwaller,~P.; Campi,~D.; Merkys,~A.; Marrazzo,~A.;
  Sohier,~T.; Castelli,~I.~E.; Cepellotti,~A.; Pizzi,~G., \latin{et~al.}
  Two-dimensional materials from high-throughput computational exfoliation of
  experimentally known compounds. \emph{Nature Nanotechnology} \textbf{2018},
  \emph{13}, 246--252\relax
\mciteBstWouldAddEndPuncttrue
\mciteSetBstMidEndSepPunct{\mcitedefaultmidpunct}
{\mcitedefaultendpunct}{\mcitedefaultseppunct}\relax
\EndOfBibitem
\bibitem[Sakuma(1988)]{sakuma1988crystal}
Sakuma,~T. Crystal structure of $\beta$-CuI. \emph{Journal of the Physical
  Society of Japan} \textbf{1988}, \emph{57}, 565--569\relax
\mciteBstWouldAddEndPuncttrue
\mciteSetBstMidEndSepPunct{\mcitedefaultmidpunct}
{\mcitedefaultendpunct}{\mcitedefaultseppunct}\relax
\EndOfBibitem
\bibitem[Keen and Hull(1994)Keen, and Hull]{keen1994determination}
Keen,~D.; Hull,~S. Determination of the structure of beta-CuI by
  high-resolution neutron powder diffraction. \emph{Journal of Physics:
  Condensed Matter} \textbf{1994}, \emph{6}, 1637\relax
\mciteBstWouldAddEndPuncttrue
\mciteSetBstMidEndSepPunct{\mcitedefaultmidpunct}
{\mcitedefaultendpunct}{\mcitedefaultseppunct}\relax
\EndOfBibitem
\bibitem[Yao \latin{et~al.}(2018)Yao, Chen, Zhang, Li, Ai, Ma, Zhao, Sun, Wu,
  Tang, \latin{et~al.} others]{yao2018synthesis}
Yao,~K.; Chen,~P.; Zhang,~Z.; Li,~J.; Ai,~R.; Ma,~H.; Zhao,~B.; Sun,~G.;
  Wu,~R.; Tang,~X., \latin{et~al.}  Synthesis of ultrathin two-dimensional
  nanosheets and van der Waals heterostructures from non-layered $\gamma$-CuI.
  \emph{npj 2D Materials and Applications} \textbf{2018}, \emph{2}, 1--7\relax
\mciteBstWouldAddEndPuncttrue
\mciteSetBstMidEndSepPunct{\mcitedefaultmidpunct}
{\mcitedefaultendpunct}{\mcitedefaultseppunct}\relax
\EndOfBibitem
\bibitem[Hayakawa \latin{et~al.}(2021)Hayakawa, Arakawa, Kono, Handa, Hayashi,
  Minamikawa, Horio, and Terasaki]{hayakawa2021x}
Hayakawa,~T.; Arakawa,~M.; Kono,~S.; Handa,~T.; Hayashi,~N.; Minamikawa,~K.;
  Horio,~T.; Terasaki,~A. X-ray absorption spectroscopy of small copper-oxide
  cluster ions for analyses of Cu oxidation state and Ar complexation: CuOAr+
  and Cu2O2+. \emph{Zeitschrift f{\"u}r Physikalische Chemie} \textbf{2021},
  \emph{235}, 213--224\relax
\mciteBstWouldAddEndPuncttrue
\mciteSetBstMidEndSepPunct{\mcitedefaultmidpunct}
{\mcitedefaultendpunct}{\mcitedefaultseppunct}\relax
\EndOfBibitem
\bibitem[Meulenberg \latin{et~al.}(2004)Meulenberg, van Buuren, Hanif, Willey,
  Strouse, and Terminello]{meulenberg2004structure}
Meulenberg,~R.~W.; van Buuren,~T.; Hanif,~K.~M.; Willey,~T.~M.; Strouse,~G.~F.;
  Terminello,~L.~J. Structure and composition of Cu-doped CdSe nanocrystals
  using soft X-ray absorption spectroscopy. \emph{Nano letters} \textbf{2004},
  \emph{4}, 2277--2285\relax
\mciteBstWouldAddEndPuncttrue
\mciteSetBstMidEndSepPunct{\mcitedefaultmidpunct}
{\mcitedefaultendpunct}{\mcitedefaultseppunct}\relax
\EndOfBibitem
\bibitem[Haigh \latin{et~al.}(2012)Haigh, Gholinia, Jalil, Romani, Britnell,
  Elias, Novoselov, Ponomarenko, Geim, and Gorbachev]{haigh2012cross}
Haigh,~S.; Gholinia,~A.; Jalil,~R.; Romani,~S.; Britnell,~L.; Elias,~D.;
  Novoselov,~K.; Ponomarenko,~L.; Geim,~A.; Gorbachev,~R. Cross-sectional
  imaging of individual layers and buried interfaces of graphene-based
  heterostructures and superlattices. \emph{Nature Materials} \textbf{2012},
  \emph{11}, 764--767\relax
\mciteBstWouldAddEndPuncttrue
\mciteSetBstMidEndSepPunct{\mcitedefaultmidpunct}
{\mcitedefaultendpunct}{\mcitedefaultseppunct}\relax
\EndOfBibitem
\bibitem[Hofer \latin{et~al.}(2021)Hofer, Skakalova, Mustonen, and
  Pennycook]{hofer2021threedimensional}
Hofer,~C.; Skakalova,~V.; Mustonen,~K.; Pennycook,~T.~J. Three-dimensional
  characterization of 2D materials using few-tilt microsecond dwell time
  electron ptychography. 2021; preprint,
  \url{https://arxiv.org/abs/2108.04625}\relax
\mciteBstWouldAddEndPuncttrue
\mciteSetBstMidEndSepPunct{\mcitedefaultmidpunct}
{\mcitedefaultendpunct}{\mcitedefaultseppunct}\relax
\EndOfBibitem
\bibitem[Hofer \latin{et~al.}(2018)Hofer, Kramberger, Monazam, Mangler,
  Mittelberger, Argentero, Kotakoski, and Meyer]{hofer2018revealing}
Hofer,~C.; Kramberger,~C.; Monazam,~M. R.~A.; Mangler,~C.; Mittelberger,~A.;
  Argentero,~G.; Kotakoski,~J.; Meyer,~J.~C. Revealing the 3D structure of
  graphene defects. \emph{2D Materials} \textbf{2018}, \emph{5}, 045029\relax
\mciteBstWouldAddEndPuncttrue
\mciteSetBstMidEndSepPunct{\mcitedefaultmidpunct}
{\mcitedefaultendpunct}{\mcitedefaultseppunct}\relax
\EndOfBibitem
\bibitem[Pennycook \latin{et~al.}(2015)Pennycook, Lupini, Yang, Murfitt, Jones,
  and Nellist]{Pennycook2015}
Pennycook,~T.~J.; Lupini,~A.~R.; Yang,~H.; Murfitt,~M.~F.; Jones,~L.;
  Nellist,~P.~D. Efficient phase contrast imaging in STEM using a pixelated
  detector. Part 1: Experimental demonstration at atomic resolution.
  \emph{Ultramicroscopy} \textbf{2015}, \emph{151}, 160--167, Special Issue:
  80th Birthday of Harald Rose; PICO 2015 – Third Conference on Frontiers of
  Aberration Corrected Electron Microscopy\relax
\mciteBstWouldAddEndPuncttrue
\mciteSetBstMidEndSepPunct{\mcitedefaultmidpunct}
{\mcitedefaultendpunct}{\mcitedefaultseppunct}\relax
\EndOfBibitem
\bibitem[Pennycook \latin{et~al.}(2019)Pennycook, Martinez, Nellist, and
  Meyer]{PENNYCOOK2019131}
Pennycook,~T.~J.; Martinez,~G.~T.; Nellist,~P.~D.; Meyer,~J.~C. High dose
  efficiency atomic resolution imaging via electron ptychography.
  \emph{Ultramicroscopy} \textbf{2019}, \emph{196}, 131--135\relax
\mciteBstWouldAddEndPuncttrue
\mciteSetBstMidEndSepPunct{\mcitedefaultmidpunct}
{\mcitedefaultendpunct}{\mcitedefaultseppunct}\relax
\EndOfBibitem
\bibitem[Yang \latin{et~al.}(2016)Yang, Rutte, Jones, Simson, Sagawa, Ryll,
  Huth, Pennycook, Green, Soltau, Kondo, Davis, and Nellist]{Yang2016}
Yang,~H.; Rutte,~R.~N.; Jones,~L.; Simson,~M.; Sagawa,~R.; Ryll,~H.; Huth,~M.;
  Pennycook,~T.~J.; Green,~M.~L.; Soltau,~H.; Kondo,~Y.; Davis,~B.~G.;
  Nellist,~P.~D. Simultaneous atomic-resolution electron ptychography and
  Z-contrast imaging of light and heavy elements in complex nanostructures.
  \emph{Nature Communications} \textbf{2016}, \emph{7}, 1--8\relax
\mciteBstWouldAddEndPuncttrue
\mciteSetBstMidEndSepPunct{\mcitedefaultmidpunct}
{\mcitedefaultendpunct}{\mcitedefaultseppunct}\relax
\EndOfBibitem
\bibitem[Huang \latin{et~al.}(2020)Huang, Yan, Zhou, Wang, Song, and
  Zhou]{huang2020group}
Huang,~X.; Yan,~L.; Zhou,~Y.; Wang,~Y.; Song,~H.-Z.; Zhou,~L. Group 11
  Transition-Metal Halide Monolayers: High Promises for Photocatalysis and
  Quantum Cutting. \emph{The Journal of Physical Chemistry Letters}
  \textbf{2020}, \emph{12}, 525--531\relax
\mciteBstWouldAddEndPuncttrue
\mciteSetBstMidEndSepPunct{\mcitedefaultmidpunct}
{\mcitedefaultendpunct}{\mcitedefaultseppunct}\relax
\EndOfBibitem
\bibitem[Klime{\v{s}} \latin{et~al.}(2009)Klime{\v{s}}, Bowler, and
  Michaelides]{klime_2009}
Klime{\v{s}},~J.; Bowler,~D.~R.; Michaelides,~A. Chemical accuracy for the van
  der Waals density functional. \emph{Journal of Physics: Condensed Matter}
  \textbf{2009}, \emph{22}, 022201\relax
\mciteBstWouldAddEndPuncttrue
\mciteSetBstMidEndSepPunct{\mcitedefaultmidpunct}
{\mcitedefaultendpunct}{\mcitedefaultseppunct}\relax
\EndOfBibitem
\bibitem[Kiselev \latin{et~al.}(2013)Kiselev, Kumskov, Zhigalina, Verbitskiy,
  Yashina, Chuvilin, Vasiliev, and Eliseev]{kiselev2013structure}
Kiselev,~N.; Kumskov,~A.; Zhigalina,~V.; Verbitskiy,~N.; Yashina,~L.;
  Chuvilin,~A.; Vasiliev,~A.; Eliseev,~A. The structure and electronic
  properties of copper iodide 1D nanocrystals within single walled carbon
  nanotubes. Journal of Physics: Conference Series. 2013; p 012035\relax
\mciteBstWouldAddEndPuncttrue
\mciteSetBstMidEndSepPunct{\mcitedefaultmidpunct}
{\mcitedefaultendpunct}{\mcitedefaultseppunct}\relax
\EndOfBibitem
\bibitem[Hull \latin{et~al.}(1998)Hull, Keen, Hayes, and
  Gardner]{hull1998superionic}
Hull,~S.; Keen,~D.; Hayes,~W.; Gardner,~N. Superionic behaviour in copper (I)
  iodide at elevated pressures and temperatures. \emph{Journal of Physics:
  Condensed Matter} \textbf{1998}, \emph{10}, 10941\relax
\mciteBstWouldAddEndPuncttrue
\mciteSetBstMidEndSepPunct{\mcitedefaultmidpunct}
{\mcitedefaultendpunct}{\mcitedefaultseppunct}\relax
\EndOfBibitem
\bibitem[Phillips(1970)]{phillips1970ionicity}
Phillips,~J. Ionicity of the chemical bond in crystals. \emph{Reviews of Modern
  Physics} \textbf{1970}, \emph{42}, 317\relax
\mciteBstWouldAddEndPuncttrue
\mciteSetBstMidEndSepPunct{\mcitedefaultmidpunct}
{\mcitedefaultendpunct}{\mcitedefaultseppunct}\relax
\EndOfBibitem
\bibitem[Ahn and Park(2016)Ahn, and Park]{ahn2016cuprous}
Ahn,~D.; Park,~S.-H. Cuprous halides semiconductors as a new means for highly
  efficient light-emitting diodes. \emph{Scientific Reports} \textbf{2016},
  \emph{6}, 1--9\relax
\mciteBstWouldAddEndPuncttrue
\mciteSetBstMidEndSepPunct{\mcitedefaultmidpunct}
{\mcitedefaultendpunct}{\mcitedefaultseppunct}\relax
\EndOfBibitem
\bibitem[Miyake \latin{et~al.}(1952)Miyake, Hoshino, and
  Takenaka]{miyake1952phase}
Miyake,~S.; Hoshino,~S.; Takenaka,~T. On the phase transition in cuprous
  iodide. \emph{Journal of the Physical Society of Japan} \textbf{1952},
  \emph{7}, 19--24\relax
\mciteBstWouldAddEndPuncttrue
\mciteSetBstMidEndSepPunct{\mcitedefaultmidpunct}
{\mcitedefaultendpunct}{\mcitedefaultseppunct}\relax
\EndOfBibitem
\bibitem[Krug and Sieg(1952)Krug, and Sieg]{KrugSieg1952}
Krug,~J.; Sieg,~L. Die Struktur der Hochtemperatur-Modifikationen des CuBr und
  CuJ. \emph{Zeitschrift für Naturforschung A} \textbf{1952}, \emph{7},
  369--371\relax
\mciteBstWouldAddEndPuncttrue
\mciteSetBstMidEndSepPunct{\mcitedefaultmidpunct}
{\mcitedefaultendpunct}{\mcitedefaultseppunct}\relax
\EndOfBibitem
\bibitem[B{\"u}hrer and H{\"a}lg(1977)B{\"u}hrer, and
  H{\"a}lg]{buhrer1977crystal}
B{\"u}hrer,~W.; H{\"a}lg,~W. Crystal structure of high-temperature cuprous
  iodide and cuprous bromide. International symposium on solid ionic and
  ionic-electronic conductors. 1977; pp 701--704\relax
\mciteBstWouldAddEndPuncttrue
\mciteSetBstMidEndSepPunct{\mcitedefaultmidpunct}
{\mcitedefaultendpunct}{\mcitedefaultseppunct}\relax
\EndOfBibitem
\bibitem[Liu \latin{et~al.}(2019)Liu, Hou, Wang, Zhang, Sun, and
  Meng]{liu2019ideal}
Liu,~J.; Hou,~W.; Wang,~E.; Zhang,~S.; Sun,~J.-T.; Meng,~S. Ideal type-II Weyl
  phonons in wurtzite CuI. \emph{Physical Review B} \textbf{2019}, \emph{100},
  081204\relax
\mciteBstWouldAddEndPuncttrue
\mciteSetBstMidEndSepPunct{\mcitedefaultmidpunct}
{\mcitedefaultendpunct}{\mcitedefaultseppunct}\relax
\EndOfBibitem
\bibitem[Hern{\'a}ndez-Cocoletzi \latin{et~al.}(2009)Hern{\'a}ndez-Cocoletzi,
  Cocoletzi, Rivas-Silva, Flores, and Takeuchi]{hernandez2009density}
Hern{\'a}ndez-Cocoletzi,~H.; Cocoletzi,~G.~H.; Rivas-Silva,~J.; Flores,~A.;
  Takeuchi,~N. Density Functional Study of the Structural Properties of Copper
  Iodide: LDA vs GGA Calculations. Journal of Nano Research. 2009; pp
  25--30\relax
\mciteBstWouldAddEndPuncttrue
\mciteSetBstMidEndSepPunct{\mcitedefaultmidpunct}
{\mcitedefaultendpunct}{\mcitedefaultseppunct}\relax
\EndOfBibitem
\bibitem[Klein \latin{et~al.}(2018)Klein, MacNeill, Lado, Soriano,
  Navarro-Moratalla, Watanabe, Taniguchi, Manni, Canfield,
  Fern{\'a}ndez-Rossier, \latin{et~al.} others]{klein2018probing}
Klein,~D.~R.; MacNeill,~D.; Lado,~J.~L.; Soriano,~D.; Navarro-Moratalla,~E.;
  Watanabe,~K.; Taniguchi,~T.; Manni,~S.; Canfield,~P.;
  Fern{\'a}ndez-Rossier,~J., \latin{et~al.}  Probing magnetism in 2D van der
  Waals crystalline insulators via electron tunneling. \emph{Science}
  \textbf{2018}, \emph{360}, 1218--1222\relax
\mciteBstWouldAddEndPuncttrue
\mciteSetBstMidEndSepPunct{\mcitedefaultmidpunct}
{\mcitedefaultendpunct}{\mcitedefaultseppunct}\relax
\EndOfBibitem
\bibitem[Huang \latin{et~al.}(2017)Huang, Clark, Navarro-Moratalla, Klein,
  Cheng, Seyler, Zhong, Schmidgall, McGuire, Cobden, \latin{et~al.}
  others]{huang2017layer}
Huang,~B.; Clark,~G.; Navarro-Moratalla,~E.; Klein,~D.~R.; Cheng,~R.;
  Seyler,~K.~L.; Zhong,~D.; Schmidgall,~E.; McGuire,~M.~A.; Cobden,~D.~H.,
  \latin{et~al.}  Layer-dependent ferromagnetism in a van der Waals crystal
  down to the monolayer limit. \emph{Nature} \textbf{2017}, \emph{546},
  270--273\relax
\mciteBstWouldAddEndPuncttrue
\mciteSetBstMidEndSepPunct{\mcitedefaultmidpunct}
{\mcitedefaultendpunct}{\mcitedefaultseppunct}\relax
\EndOfBibitem
\bibitem[Kashin \latin{et~al.}(2020)Kashin, Mazurenko, Katsnelson, and
  Rudenko]{kashin2020orbitally}
Kashin,~I.; Mazurenko,~V.; Katsnelson,~M.; Rudenko,~A. Orbitally-resolved
  ferromagnetism of monolayer CrI3. \emph{2D Materials} \textbf{2020},
  \emph{7}, 025036\relax
\mciteBstWouldAddEndPuncttrue
\mciteSetBstMidEndSepPunct{\mcitedefaultmidpunct}
{\mcitedefaultendpunct}{\mcitedefaultseppunct}\relax
\EndOfBibitem
\bibitem[Wang and Sanyal(2021)Wang, and Sanyal]{wang2021systematic}
Wang,~D.; Sanyal,~B. Systematic Study of Monolayer to Trilayer CrI3: Stacking
  Sequence Dependence of Electronic Structure and Magnetism. \emph{The Journal
  of Physical Chemistry C,} \textbf{2021}, Advance online publication, doi:
  10.1021/acs.jpcc.1c04311\relax
\mciteBstWouldAddEndPuncttrue
\mciteSetBstMidEndSepPunct{\mcitedefaultmidpunct}
{\mcitedefaultendpunct}{\mcitedefaultseppunct}\relax
\EndOfBibitem
\bibitem[Amoroso \latin{et~al.}(2020)Amoroso, Barone, and
  Picozzi]{amoroso2020spontaneous}
Amoroso,~D.; Barone,~P.; Picozzi,~S. Spontaneous skyrmionic lattice from
  anisotropic symmetric exchange in a Ni-halide monolayer. \emph{Nature
  communications} \textbf{2020}, \emph{11}, 1--9\relax
\mciteBstWouldAddEndPuncttrue
\mciteSetBstMidEndSepPunct{\mcitedefaultmidpunct}
{\mcitedefaultendpunct}{\mcitedefaultseppunct}\relax
\EndOfBibitem
\bibitem[Kurumaji \latin{et~al.}(2013)Kurumaji, Seki, Ishiwata, Murakawa,
  Kaneko, and Tokura]{kurumaji2013magnetoelectric}
Kurumaji,~T.; Seki,~S.; Ishiwata,~S.; Murakawa,~H.; Kaneko,~Y.; Tokura,~Y.
  Magnetoelectric responses induced by domain rearrangement and spin structural
  change in triangular-lattice helimagnets NiI 2 and CoI 2. \emph{Physical
  Review B} \textbf{2013}, \emph{87}, 014429\relax
\mciteBstWouldAddEndPuncttrue
\mciteSetBstMidEndSepPunct{\mcitedefaultmidpunct}
{\mcitedefaultendpunct}{\mcitedefaultseppunct}\relax
\EndOfBibitem
\bibitem[Eigler and Hirsch(2014)Eigler, and Hirsch]{eigler2014chemistry}
Eigler,~S.; Hirsch,~A. Chemistry with graphene and graphene oxide—challenges
  for synthetic chemists. \emph{Angewandte Chemie International Edition}
  \textbf{2014}, \emph{53}, 7720--7738\relax
\mciteBstWouldAddEndPuncttrue
\mciteSetBstMidEndSepPunct{\mcitedefaultmidpunct}
{\mcitedefaultendpunct}{\mcitedefaultseppunct}\relax
\EndOfBibitem
\bibitem[Poikela \latin{et~al.}(2014)Poikela, Plosila, Westerlund, Campbell,
  De~Gaspari, Llopart, Gromov, Kluit, Van~Beuzekom, Zappon, \latin{et~al.}
  others]{poikela2014timepix3}
Poikela,~T.; Plosila,~J.; Westerlund,~T.; Campbell,~M.; De~Gaspari,~M.;
  Llopart,~X.; Gromov,~V.; Kluit,~R.; Van~Beuzekom,~M.; Zappon,~F.,
  \latin{et~al.}  Timepix3: a 65K channel hybrid pixel readout chip with
  simultaneous ToA/ToT and sparse readout. \emph{Journal of Instrumentation}
  \textbf{2014}, \emph{9}, C05013\relax
\mciteBstWouldAddEndPuncttrue
\mciteSetBstMidEndSepPunct{\mcitedefaultmidpunct}
{\mcitedefaultendpunct}{\mcitedefaultseppunct}\relax
\EndOfBibitem
\bibitem[Jannis \latin{et~al.}(2021)Jannis, Hofer, Gao, Xie, Béché,
  Pennycook, and Verbeeck]{jannis2021event}
Jannis,~D.; Hofer,~C.; Gao,~C.; Xie,~X.; Béché,~A.; Pennycook,~T.~J.;
  Verbeeck,~J. Event driven 4D STEM acquisition with a Timepix3 detector:
  microsecond dwell time and faster scans for high precision and low dose
  applications. 2021\relax
\mciteBstWouldAddEndPuncttrue
\mciteSetBstMidEndSepPunct{\mcitedefaultmidpunct}
{\mcitedefaultendpunct}{\mcitedefaultseppunct}\relax
\EndOfBibitem
\bibitem[Susi \latin{et~al.}(2019)Susi, Madsen, Ludacka, Mortensen, Pennycook,
  Lee, Kotakoski, Kaiser, and Meyer]{SUSI2019Efficient}
Susi,~T.; Madsen,~J.; Ludacka,~U.; Mortensen,~J.~J.; Pennycook,~T.~J.; Lee,~Z.;
  Kotakoski,~J.; Kaiser,~U.; Meyer,~J.~C. Efficient first principles simulation
  of electron scattering factors for transmission electron microscopy.
  \emph{Ultramicroscopy} \textbf{2019}, \emph{197}, 16--22\relax
\mciteBstWouldAddEndPuncttrue
\mciteSetBstMidEndSepPunct{\mcitedefaultmidpunct}
{\mcitedefaultendpunct}{\mcitedefaultseppunct}\relax
\EndOfBibitem
\bibitem[Enkovaara \latin{et~al.}(2010)Enkovaara, Rostgaard, Mortensen, Chen,
  Dulak, Ferrighi, Gavnholt, Glinsvad, Haikola, Hansen, Kristoffersen, Kuisma,
  Larsen, Lehtovaara, Ljungberg, Lopez-Acevedo, Moses, Ojanen, Olsen, Petzold,
  Romero, Stausholm-Møller, Strange, Tritsaris, Vanin, Walter, Hammer,
  Häkkinen, Madsen, Nieminen, Nørskov, Puska, Rantala, Schiøtz, Thygesen,
  and Jacobsen]{enkovaara_electronic_2010}
Enkovaara,~J. \latin{et~al.}  Electronic structure calculations with {GPAW}: a
  real-space implementation of the projector augmented-wave method.
  \emph{Journal of Physics: Condensed Matter} \textbf{2010}, \emph{22},
  253202\relax
\mciteBstWouldAddEndPuncttrue
\mciteSetBstMidEndSepPunct{\mcitedefaultmidpunct}
{\mcitedefaultendpunct}{\mcitedefaultseppunct}\relax
\EndOfBibitem
\bibitem[Bl\"ochl(1994)]{blochl_paw_1994}
Bl\"ochl,~P.~E. Projector augmented-wave method. \emph{Physical Review B}
  \textbf{1994}, \emph{50}, 17953--17979\relax
\mciteBstWouldAddEndPuncttrue
\mciteSetBstMidEndSepPunct{\mcitedefaultmidpunct}
{\mcitedefaultendpunct}{\mcitedefaultseppunct}\relax
\EndOfBibitem
\bibitem[Monkhorst and Pack(1976)Monkhorst, and Pack]{monkhorst_special_1976}
Monkhorst,~H.~J.; Pack,~J.~D. Special points for {Brillouin}-zone integrations.
  \emph{Physical Review B} \textbf{1976}, \emph{13}, 5188--5192\relax
\mciteBstWouldAddEndPuncttrue
\mciteSetBstMidEndSepPunct{\mcitedefaultmidpunct}
{\mcitedefaultendpunct}{\mcitedefaultseppunct}\relax
\EndOfBibitem
\bibitem[Krukau \latin{et~al.}(2006)Krukau, Vydrov, Izmaylov, and
  Scuseria]{krukau_2006}
Krukau,~A.~V.; Vydrov,~O.~A.; Izmaylov,~A.~F.; Scuseria,~G.~E. Influence of the
  exchange screening parameter on the performance of screened hybrid
  functionals. \emph{The Journal of Chemical Physics} \textbf{2006},
  \emph{125}, 224106\relax
\mciteBstWouldAddEndPuncttrue
\mciteSetBstMidEndSepPunct{\mcitedefaultmidpunct}
{\mcitedefaultendpunct}{\mcitedefaultseppunct}\relax
\EndOfBibitem
\end{mcitethebibliography}

\newpage

\section{Methods}

\subsection{Sample preparation}

The synthetic route to 2D h-CuI confined between layers of graphene based on a wet chemical process was developed by Danubia NanoTech s.r.o. in Bratislava. First, graphene oxide (GO) was produced by the modified Hummers' method~\cite{eigler2014chemistry} resulting in 95\% of single atom thick GO flakes with a lateral size 5-25 µm. Then 10~mL water dispersion of GO  at a concentration of 1.5~mg/mL was vacuum-filtrated through a polycarbonate (PC) membrane. Then, a cooled solution (0\textdegree\,C) of 20~mg copper chloride (CuCl$_2$) in 2~mL water was added. The last step was reduction of GO adding 1~mL of hydroiodic acid (HI). HI plays a double role: (1) H$^{+}$ reduces the oxide groups from GO forming H$_{2}$O whereas (2) I$^{-}$ anions react with Cu$^{2+}$ cations forming the novel 2D crystal structure, h-CuI. The crucial aspect is timing of the processes: CuI is formed during reduction when the interlayer distance of GO flakes ~0.8 nm collapses to 0.35 nm, the interlayer distance of the rGO planes; consequently, graphene sheets tightly wrap the 2D h-CuI crystals. The applied pressure of van der Waals force keeps the 2D h-CuI stabilized. The resulting layer was rinsed by ethanol, separated from the PC membrane and dried. Even though, between the graphene layers exclusively 2D CuI is observed, it is difficult to avoid the formation of 3D CuI on the outer (farther from filter) surface of rGO film. Therefore, in all experiments, the surface of the filtrated film was removed by exfoliation with adhesive tape. For electron microscopy, reduced graphene oxide flakes produced by the chemical process were exfoliated under an optical microscope with the micromechanical cleavage method using adhesive tape. Flakes of a few layers in thickness were adhered onto a TEM grid by pressing the exfoliated material attached on the tape against the grid. Standard Au TEM grids with 300 mesh and Quantifoil$^\mathrm{TM}$ amorphous carbon support were used. For some samples, to increase mechanical stability of small flakes, same grids with pre-transferred monolayer graphene were used.

\subsection{(Scanning) transmission electron microscopy}

Scanning transmission electron microscopy experiments were conducted using an aberration corrected Nion UltraSTEM~100 microscope in Vienna, operated at 60\,keV electron energy, with the sample in ultra-high vacuum conditions at a pressure of 10$^{-10}$~mbar. The angular range of the high-angle annular dark field detector (HAADF) used for the image acquisition was 80-300~mrad. The electron energy loss spectra were recorded with a combination of Gatan PEELS 666 spectrometer and an Andor iXon 897 electron-multiplying charge-coupled device (EMCCD) camera with the same instrument. The energy dispersion used in the experiments was 0.5~eV/pixel. The spectral maps in Figure~\ref{fig:eels} were acquired with a pixel dwell time of 150~ms and comprise of $32 \times 32$ pixel arrays, whereas the maps in Figure~\ref{fig:Cross-section} comprise of $64 \times 64$ pixel arrays. The integrated core-loss peak intensities for the spectral maps were produced by fitting a power-law background for each peak separately and integrating over the given energy range.

The nano-beam and parallel beam electron diffraction experiments were conducted using an image-corrected JEOL ARM 200F HR-TEM operated at 80~keV electron energy. The nano-beam diffraction data was acquired using a probe size of 5--10~nm with beam convergence of ca. 2~mrad, allowing a high spatial selectivity.

Microsecond dwell time electron ptychography was conducted using a fast, event-driven timepix3~\cite{poikela2014timepix3, jannis2021event} camera in a probe-corrected FEI ThemisZ instrument with a probe convergence angle of 30~mrad and a beam current of ca. 1~pA. Diffraction patterns as a function of probe position (also referred to as 4D-STEM) were simultaneously with the HAADF signal.

\subsection{TEM / STEM / electron diffraction simulations }
STEM image simulation were carried out with the multislice PyQSTEM Package~\cite{SUSI2019Efficient}. The simulation parameters were chosen to be the same as in the experimental data and the source size was chosen in order to have a similar broadening as in the experimental images. The number of slices is four for the normal incident electron beam and is increased to 30 when the model is tilted. Diffraction patterns were simulated by first carrying out a TEM simulation and by squaring the exit wave in the diffraction plane.

\subsection{X-ray absorption spectroscopy}
The absorption spectrum of the Cu $L_{2/3}$-edge was recorded at the BACH beam-line of ELETTRA synchrotron light source in the total-electron-yield mode, where a drain current was measured from the sample fixed on a carbon adhesive tape. For the accurate determination of the photon energy the photo emission spectrum Au(4f) was acquired from a gold-foil fixed on the sample plate. The BACH beamline works in the extreme UV-soft X-ray photon energy range (35-1650 eV) with selectable light polarization, high energy resolution, and high intensity and brilliance. The sample environment is completely in ultra-high-vacuum.

\subsection{3D reconstruction}
The complete 4D data set was processed to extract the phase of the electron wave by single side-band ptychography. The resulting phase images were used to increase the amount of information in the tilted projections to facilitate the visualization of lighter elements~\cite{Yang2016}. The 3D reconstruction method is based on an optimization process where the SSB and ADF simulations of a model are matched to the whole experimental data set, including all different tilt angles. Details of the theoretical framework, discussions of the accuracy and comparison with only ADF-based reconstructions can be found in Ref. \citenum{hofer2021threedimensional}.

\subsection{Density functional theory calculations}
All atomistic simulations were performed using density functional theory (DFT) as implemented in the GPAW package~\cite{enkovaara_electronic_2010}. The core electrons were described using projector augmented wave (PAW) potentials~\cite{blochl_paw_1994}. The wave functions were expanded in a plane-wave basis set with an energy cutoff of 650~eV. We used the optPBE-vdW density functional~\cite{klime_2009} to describe exchange and correlation as well as van der Waals interactions. Sampling of the Brillouin zone was performed according to the Monkhorst-Pack scheme~\cite{monkhorst_special_1976} using $\Gamma$-centred $12\times12\times9$ and $12\times12\times1$ \textit{k}-point grids for unit cells of bulk and single layer $\beta$-CuI, respectively. For the unit cell of the h-CuI/graphene heterostructure, that has a size of $3\times3$ supercell of graphene, $9\times9\times1$ \textit{k}-point grid was used. The h-CuI/graphene unit cell is shown in Figure~\ref{fig:gr_cui}. A vacuum of 20~{\AA} was included to separate slabs in the $z$-direction. The structures were optimised until the force on any atom was less than 10~meV/{\AA}$^2$. Due to the small incommunsurability between the graphene and h-CuI lattices, a strain of ca. 1.6\%~was applied to the h-CuI crystal in the calculations that included graphene, expanding the lattice constant of h-CuI from 4.20~{\AA} to 4.27~{\AA} and allowing periodic boundary conditions. To calculate the electronic band gap of CuI the Heyd–Scuseria–Ernzerhof (HSE06) hybrid DFT functional~\cite{krukau_2006} was used. 

\section{Author contributions}
P.K., V.S. and M.H. developed the method and synthesized the 2D h-CuI films
encapsulated between graphene layers. K.M. prepared the samples for electron
microscopy and conducted the STEM, whereas C.H. and J.C.M performed the HR-TEM and nano-area
electron diffraction experiments. C.H. and T.J.P. used SSB ptychography method for 3D analysis of the 2D h-CuI
structure. The DFT calculations were done by A.M. The results were analyzed by K.M., C.H.,
A.M. and V.S. with input from J.K., T.J.P., J.C.M and T.S. The manuscript was
written by K.M., C.H., J.K., T.J.P. and V.S. with contributions from all co-authors.

\begin{acknowledgement}
We acknowledge funding from the European Research Council (ERC) under the European Union’s Horizon 2020 research and innovation programme Grant agreements No.~756277-ATMEN (A.M. and T.S.) and No.~802123-HDEM (C.H. and T.J.P.). Computational resources from the Vienna Scientific Cluster (VSC) are gratefully acknowledged. V.S. was supported by the Austrian Science Fund (FWF) (project no. I2344-N36), the Slovak Research and Development Agency (APVV-16-0319), the project CEMEA of the Slovak Academy of Sciences, ITMS project code 313021T081 of the Research \& Innovation Operational Programme and from the V4-Japan Joint Research Program (BGapEng). J.K. acknowledges the FWF funding within project P31605-N36 and M.H. the funding from Slovak Research and Development Agency via the APVV-15-0693 and APVV-19-0365 project grants. Danubia NanoTech s.r.o. has received funding from the European Union’s Horizon 2020 research and innovation programme under grant agreement No 101008099 (CompSafeNano project) and also thanks Mr. Kamil Bern\'ath for his support.
\end{acknowledgement}

\begin{suppinfo}
Images of interconnected crystal domains; nano-beam electron diffraction pattern; graphene bilayer moiré at the h-CuI crystal edge; images of the flake used in the cross-sectional imaging experiment; additional details of the tilt--experiments; dynamics of a h-CuI crystal under electron beam; DFT band structure results, and the calculated binding energies for 2D h-CuI/graphene heterostructure and bulk $\beta$-CuI; STEM HAADF images of 2D AgI and NiI$_2$ in graphene encapsulation.

\appendix
\renewcommand{\thefigure}{S\arabic{figure}}
\setcounter{figure}{0}
\renewcommand{\thetable}{S\arabic{table}}
\setcounter{table}{0}

\clearpage
\newpage

\section{Supporting Information}

\subsection{Interconnected crystal domains}

\begin{figure}[ht!]
	\includegraphics[width=1.0\textwidth,keepaspectratio]{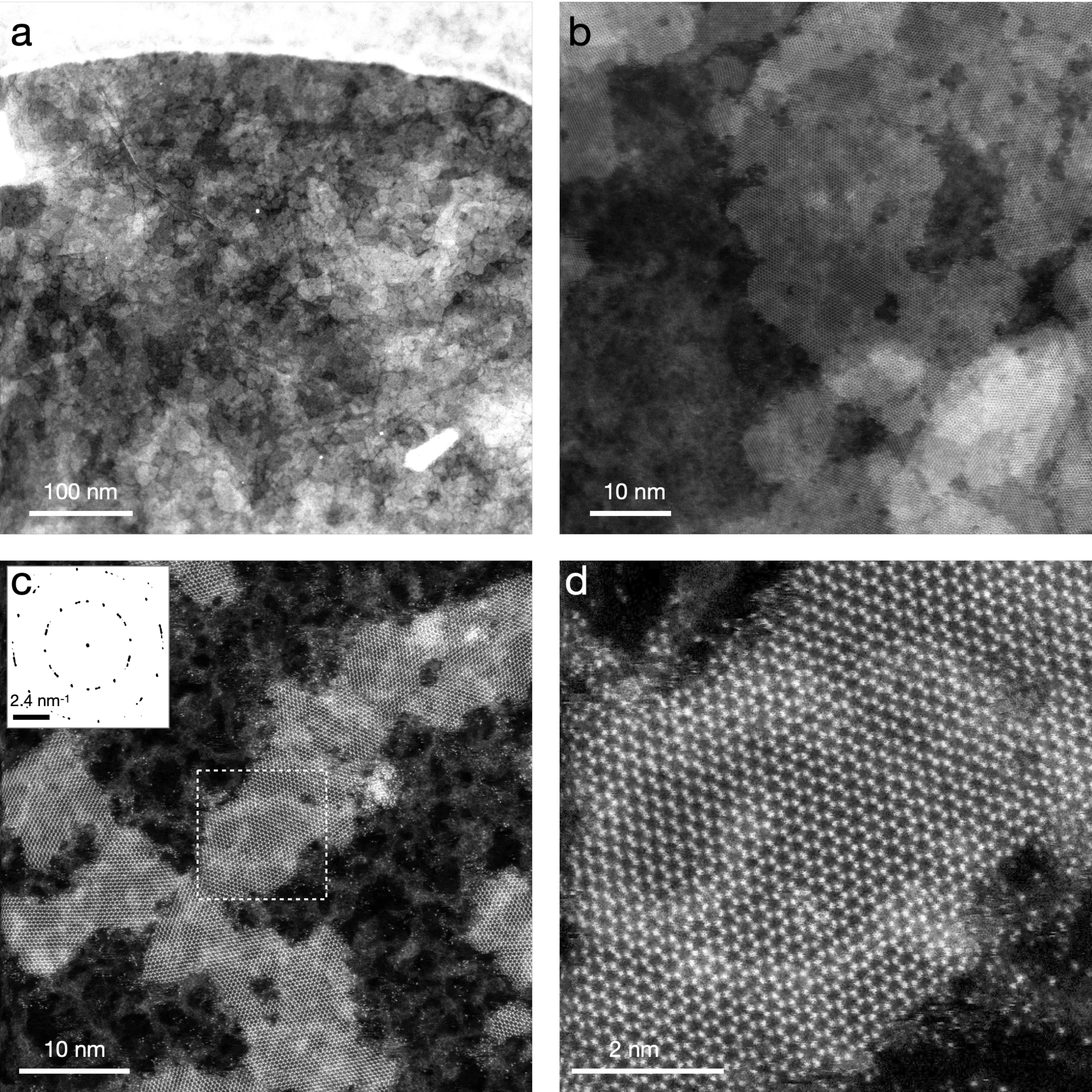}
    \caption{(a-b) Interconnected 2D h-CuI crystal domains in few-layer rGO. (c) 48 nm $\times$ 48 nm overview of interconnected domains and a Fourier transform of the image in the inset showing their respective orientations. The rectangle highlights the area from which the real-space ADF STEM tilt-series shown in Figure 4 was acquired. The highlighted area is also magnified in panel (d).}
    \label{fig:domains}
\end{figure}

\newpage

\subsection{Anisotropy in 2D h-CuI in-plane lattice constant} 

\begin{figure}[ht!]
	\includegraphics[width=0.7\textwidth,keepaspectratio]{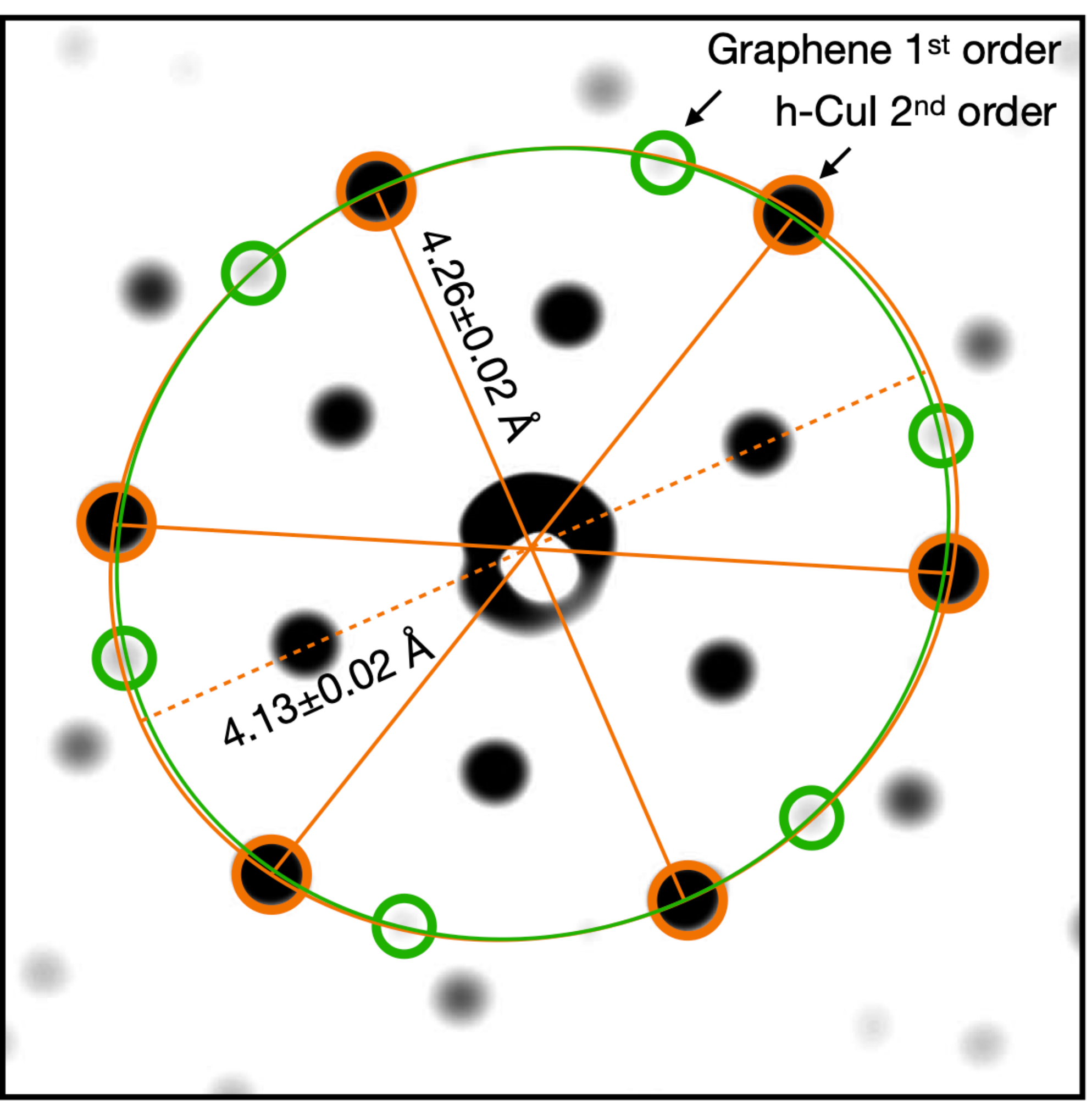}
    \caption{Nano-beam diffraction pattern of a single h-CuI crystal. The graphene 1st order and h-CuI 2nd order Bragg reflections are separately fitted with an elliptical line shape (the patterns are elliptically distorted due to a spurious sample tilt), which allows a direct comparison of the commensurability of the two structures in different directions. The measured lattice parameter values for h-CuI are indicated along the respective directions in the reciprocal space. Note that the lines are drawn only to guide the eye and thus also the $d$-spacing values tabulated next to them are indicative of distances in real-space in certain directions, and not in the reciprocal space.}
    \label{fig:diffraction}
\end{figure}

\newpage

\subsection{2D h-CuI crystal orientation with respect to graphene}

\begin{figure}[ht!]
	\includegraphics[width=1.0\textwidth,keepaspectratio]{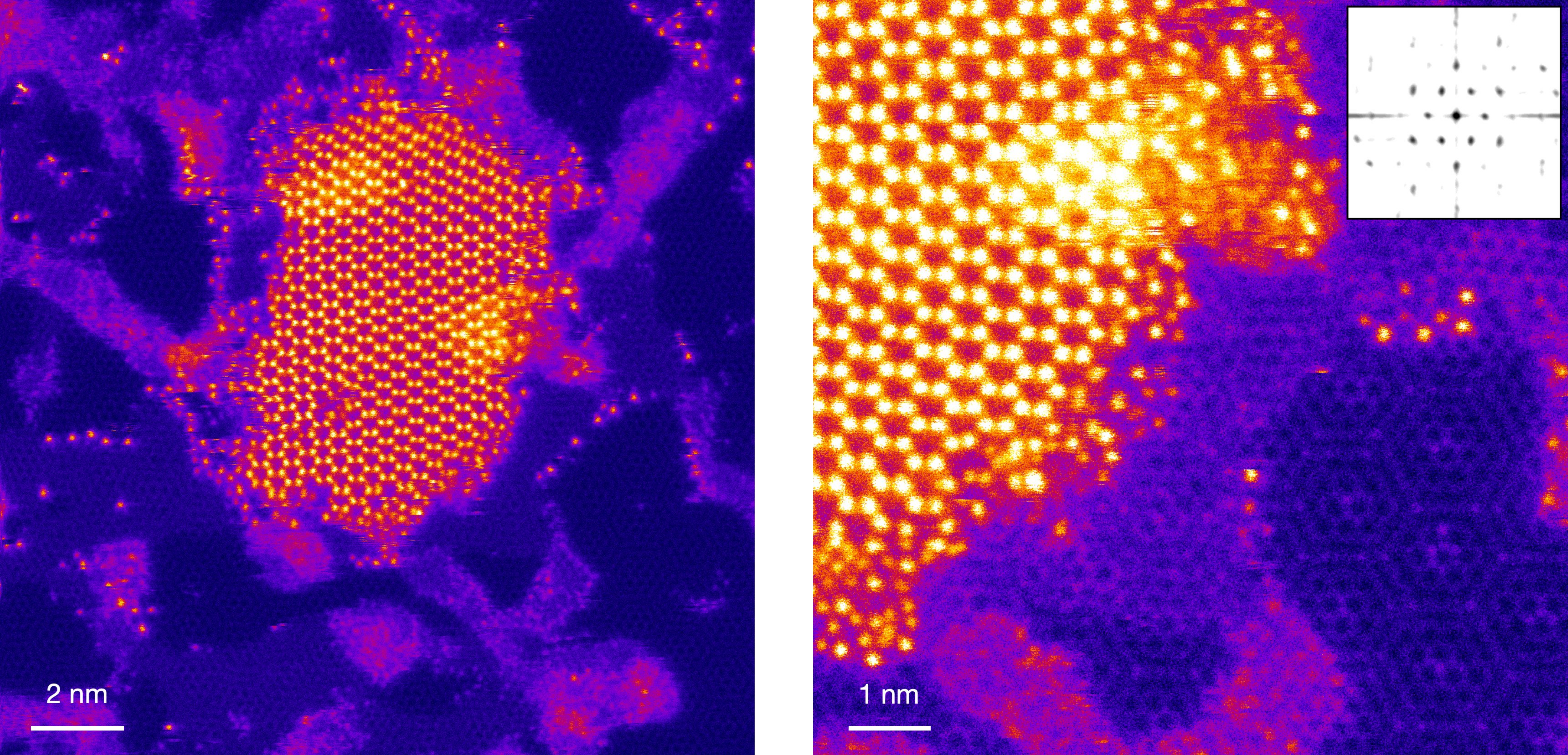}
    \caption{ADF STEM image of a small 2D h-CuI crystal and its edge. Note that the bilayer graphene moiré is visible in the background in the right panel. A Fourier transform of the right panel is shown in the inset.}
    \label{fig:moire}
\end{figure}

\newpage

\subsection{Cross-sectional imaging of 2D h-CuI}

\begin{figure}[ht!]
	\includegraphics[width=1.0\textwidth,keepaspectratio]{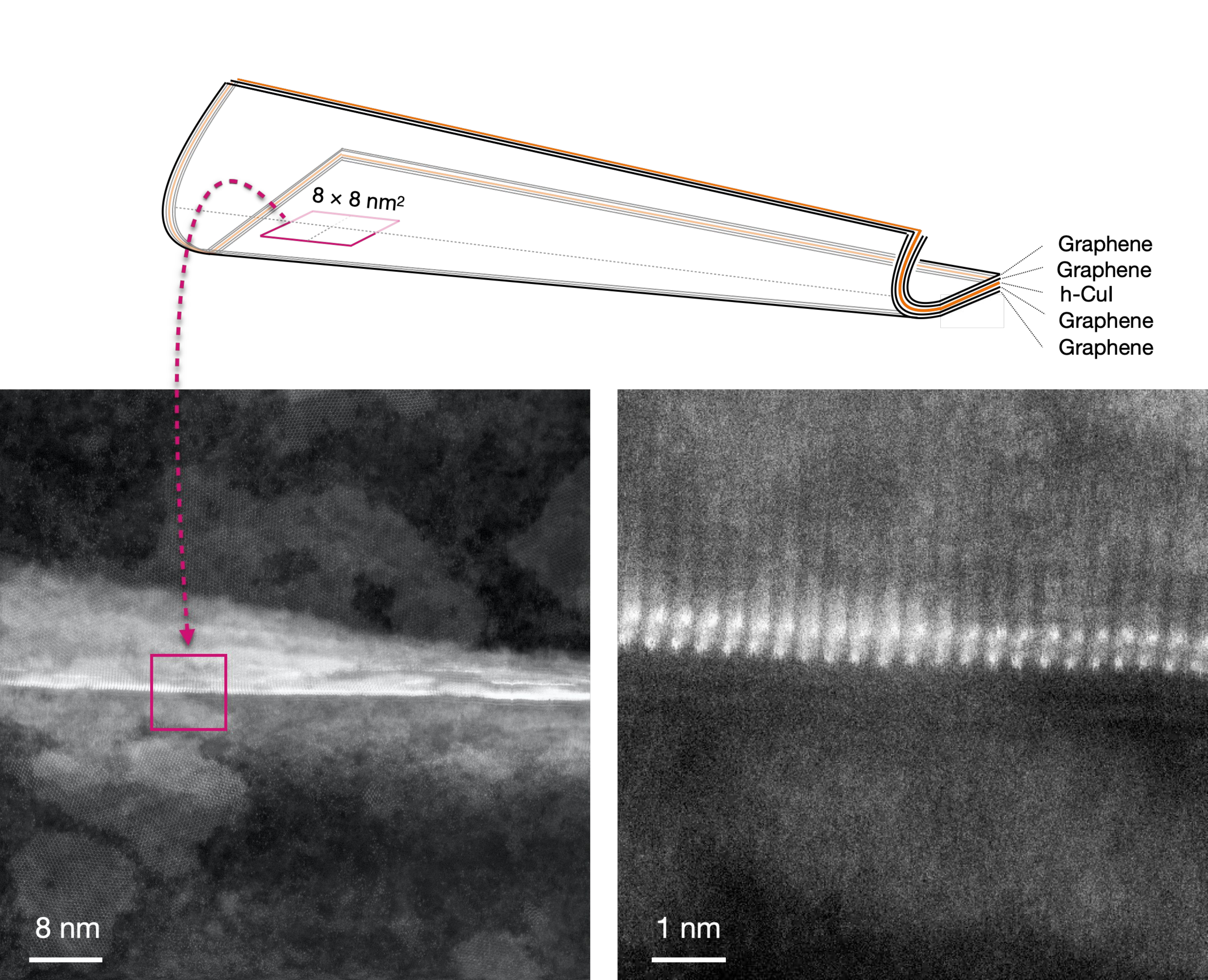}
    \caption{The sample area from which the cross-sectional images shown in Figure 3 were acquired. The top row shows a schematic drawing of the folded multilayer graphene/h-CuI heterostructure.}
    \label{fig:bent_flake}
\end{figure}

\newpage

\subsection{STEM-- and electron diffraction--tilt experiments}

\begin{figure}[ht!]
	\includegraphics[width=0.85\textwidth,keepaspectratio]{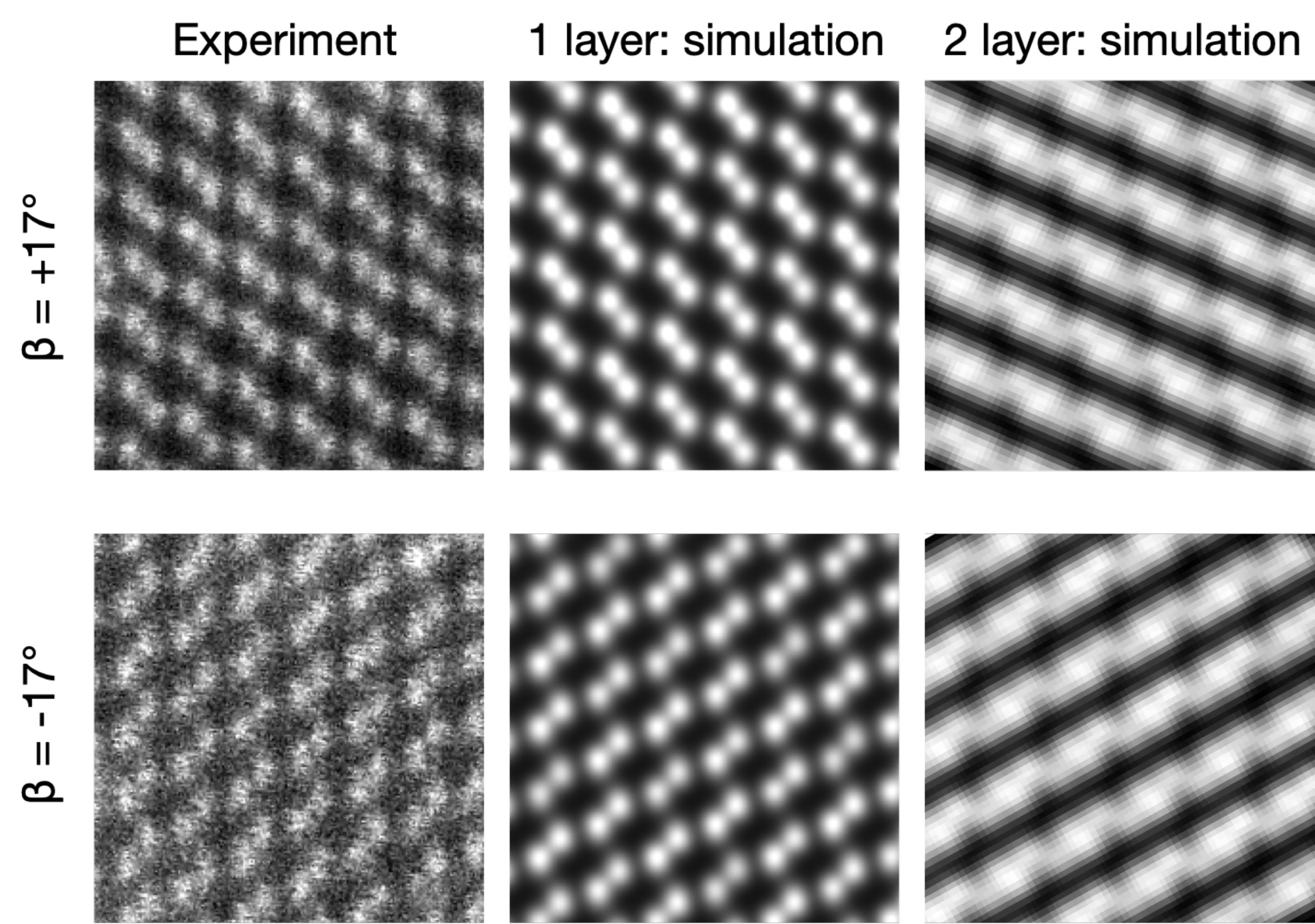}
    \caption{Experimental image of monolayer CuI (left), simulation of monolayer CuI (center) and bilayer CuI (right). The tilt angle is +17\textdegree~for the top row and -17\textdegree~for the bottom row images with a horizontal tilt axis with respect to the image plane.}
    \label{fig:tilt_SI}
\end{figure}

\begin{figure}[ht!]
	\includegraphics[width=0.85\textwidth,keepaspectratio]{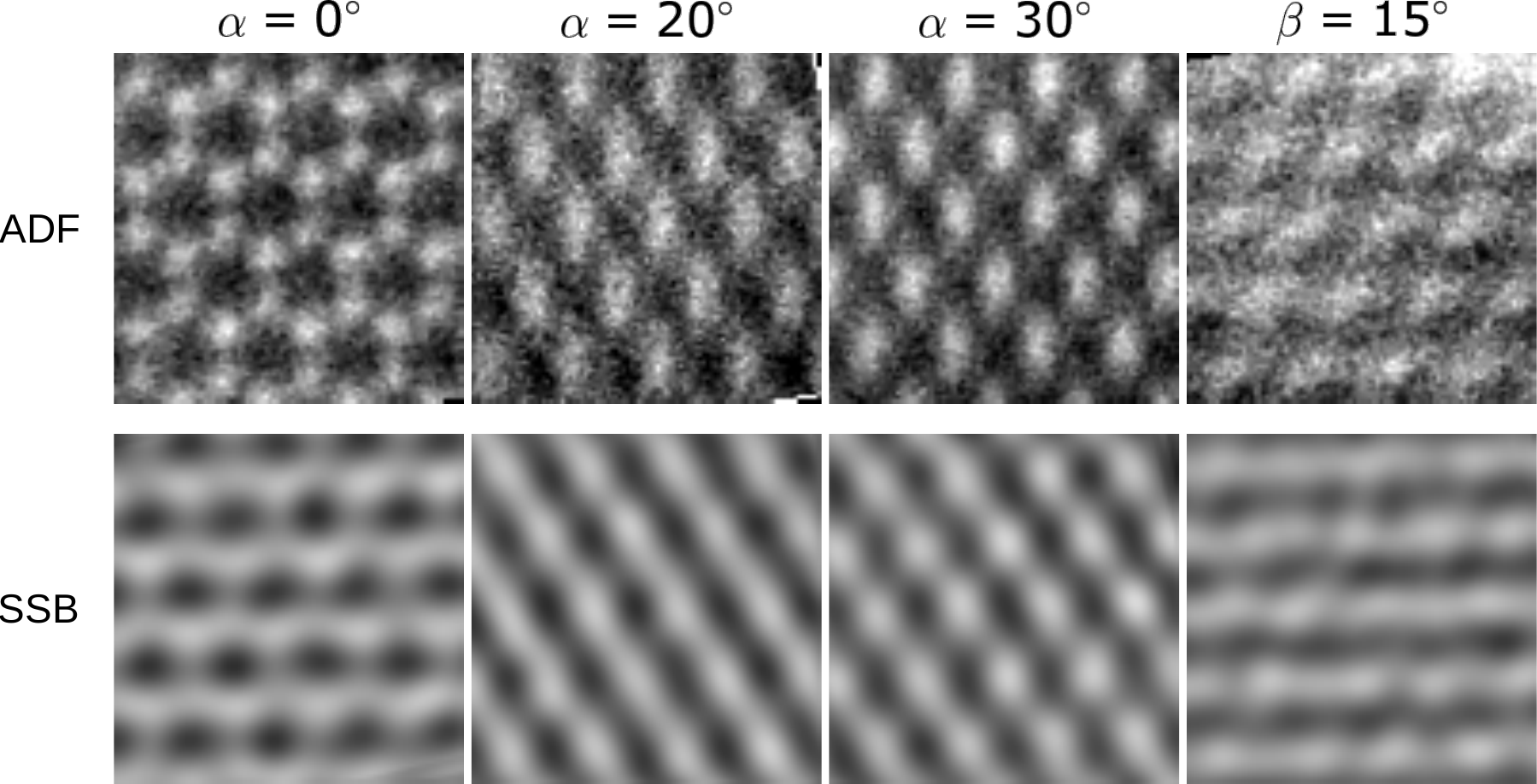}
    \caption{Few-tilt series of h-CuI ADF STEM and reconstructed SSB images used for the 3D reconstructed model in Figure 5 of the main text.}
    \label{fig:SSB}
\end{figure}

As a further method of confirming the correct identification of $\beta$-CuI we probed the diffraction space by nano-beam (NBED) electron diffraction at different tilt angles and compared the experimental intensity of the first order diffraction spots with simulated intensities. For simplicity, we only considered those orientations that are aligned with the tilt axis. As a result of the three-fold symmetry of the crystal, there is a slight break in the inversion symmetry of the diffraction peaks. This asymmetry is also observed in hexagonal boron nitride~\cite{SUSI2019Efficient}. The simulated diffraction intensities and experimental intensities (solid line and markers, respectively), are shown in Figure~\ref{fig:DPtomo} with the tilt axis vertically aligned.  Although all six peaks match well with the calculated intensities, small deviations can be explained by the finite accuracy of the tilt angle and the tilt of the structure itself. These mismatches appear more likely at high gradients of the simulated curve (e.g.\ close to 0\textdegree).

\begin{figure}[ht!]
	\includegraphics[width=1\textwidth,keepaspectratio]{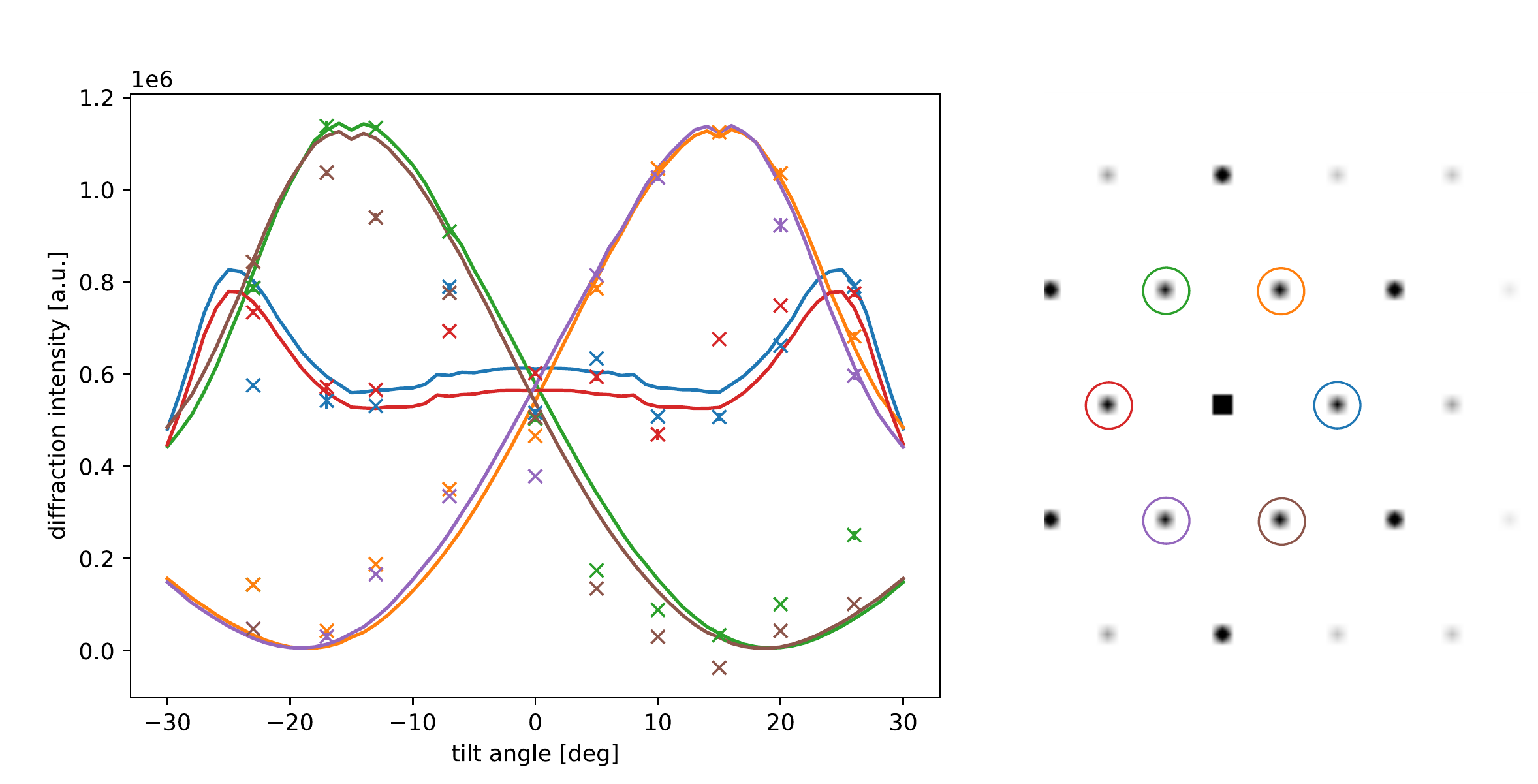}
    \caption{Simulated (solid line) and experimental (crosses) diffraction intensities of all six 1st order Bragg reflections of h-CuI as a function of tilt angle.}
    \label{fig:DPtomo}
\end{figure}

\subsection{Dynamics and stability under electron exposure}

HR-TEM image sequences were acquired at 80 keV electron energy to study the dynamics of CuI under the electron beam which gives deeper insights into the overall stability of the structure. Figure~\ref{fig:rotating} shows a three frame sequence under the optimal focus providing the ideal contrast of graphene and h-CuI. The time-resolved sequences demonstrate anisotropic degradation along the zig-zag direction, which can be explained by the release of Cu and I in the different sublattice sites at the crystal edge. While the released I atoms are trapped between the graphene layers and randomly move around, the Cu atoms diffuse outside and react with the residual oxygen gas in the TEM vacuum, forming CuO nanoparticles. The composition of these particles is confirmed by their lattice spacing and elemental composition (cf.~Figure~\ref{fig:cuo}). Further on, the size of the particles increases along with the greater electron dose, which is a result of the dissociation of further Cu atoms. The blue rectangle in Figure~\ref{fig:rotating} indicates the increasing lateral size of the particles. 

In this particular case the 2D h-CuI crystal is in alignment with the graphene sheets, as is indicated by the Fourier transforms below the frames, and small rotations around the ideal alignment are observed. However, also much larger rotations that correspond to invariant translations within the hexagonally symmetric space are possible. The structure observed here, for instance, as soon as the crystal has achieved small enough size, rotates by ca. 30\textdegree~in the last frame of the series. These discrete orientations of 2D h-CuI with respect to the graphene sheets with small deviations further elaborate the importance of the interaction between the two lattices.

\begin{figure}[!]
	\includegraphics[width=1.0\textwidth,keepaspectratio]{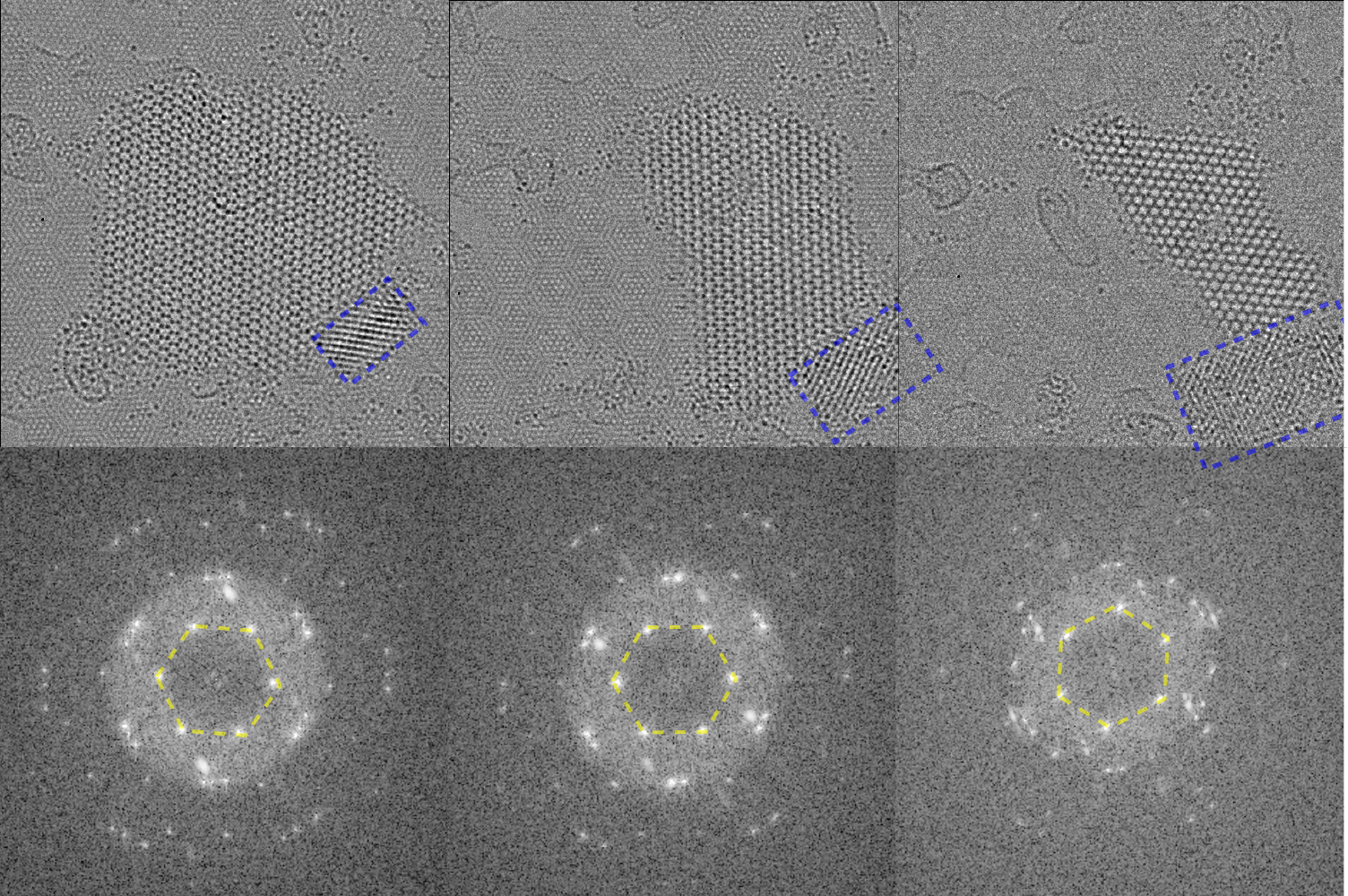}
    \caption{Series of HR-TEM images of h-CuI. The blue rectangle highlights the CuO particle that grows larger frame-by-frame. The images on the bottom row show the corresponding Fourier transforms, highlighting the orientation of h-CuI (yellow broken line).}
    \label{fig:rotating}
\end{figure}

\begin{figure}[!]
	\includegraphics[width=1.0\textwidth,keepaspectratio]{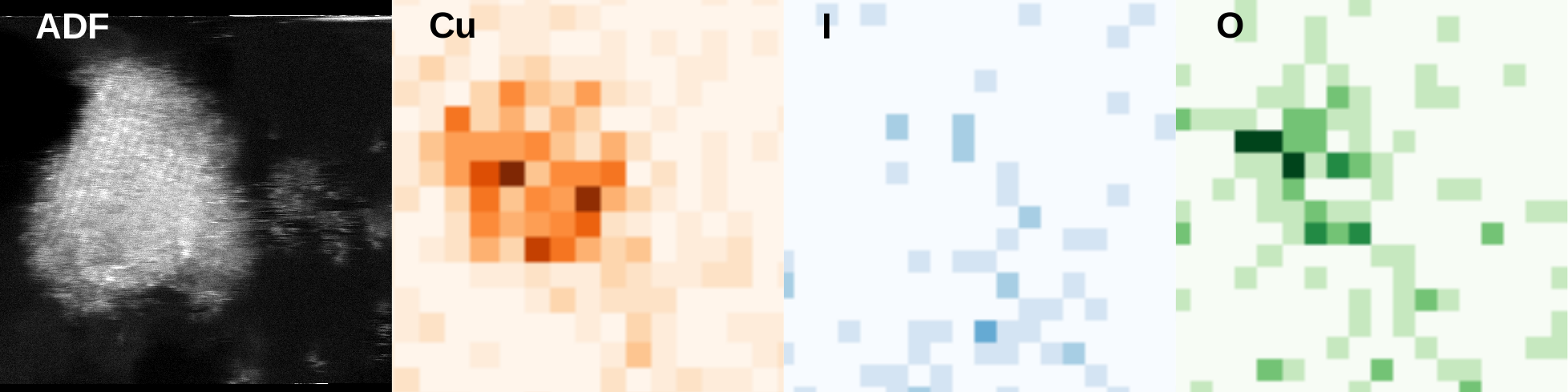}
    \caption{ADF STEM image and energy dispersive x-ray spectroscopy maps of the CuO particle formed after electron irradiation of h-CuI crystal shown in Figure~\ref{fig:rotating}.}
    \label{fig:cuo}
\end{figure}

\newpage
\subsection{Density functional theory results}

\begin{figure}[!]
	\includegraphics[width=0.8\textwidth,keepaspectratio]{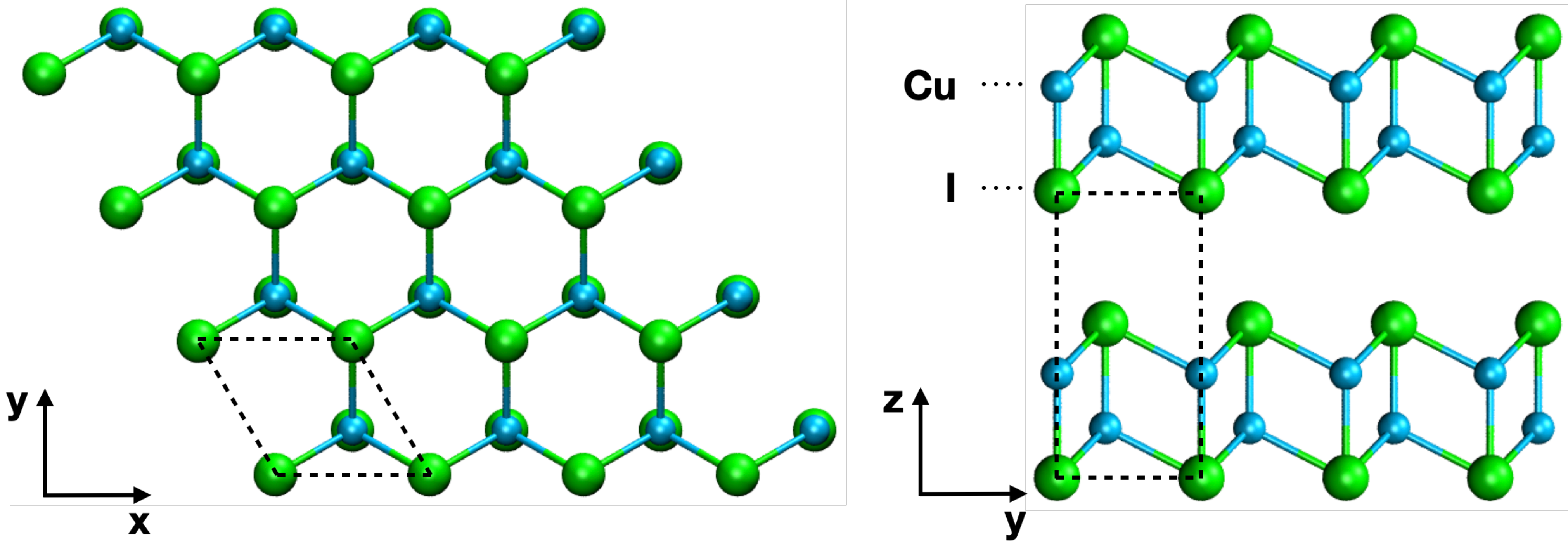}
    \caption{Atomic structure of $\beta$-CuI represented by the $4\times4\times2$ supercell with the primitive cell outlined.}
    \label{fig:bulk_cui}
\end{figure}

\begin{figure}[!]
	\includegraphics[width=0.7\textwidth,keepaspectratio]{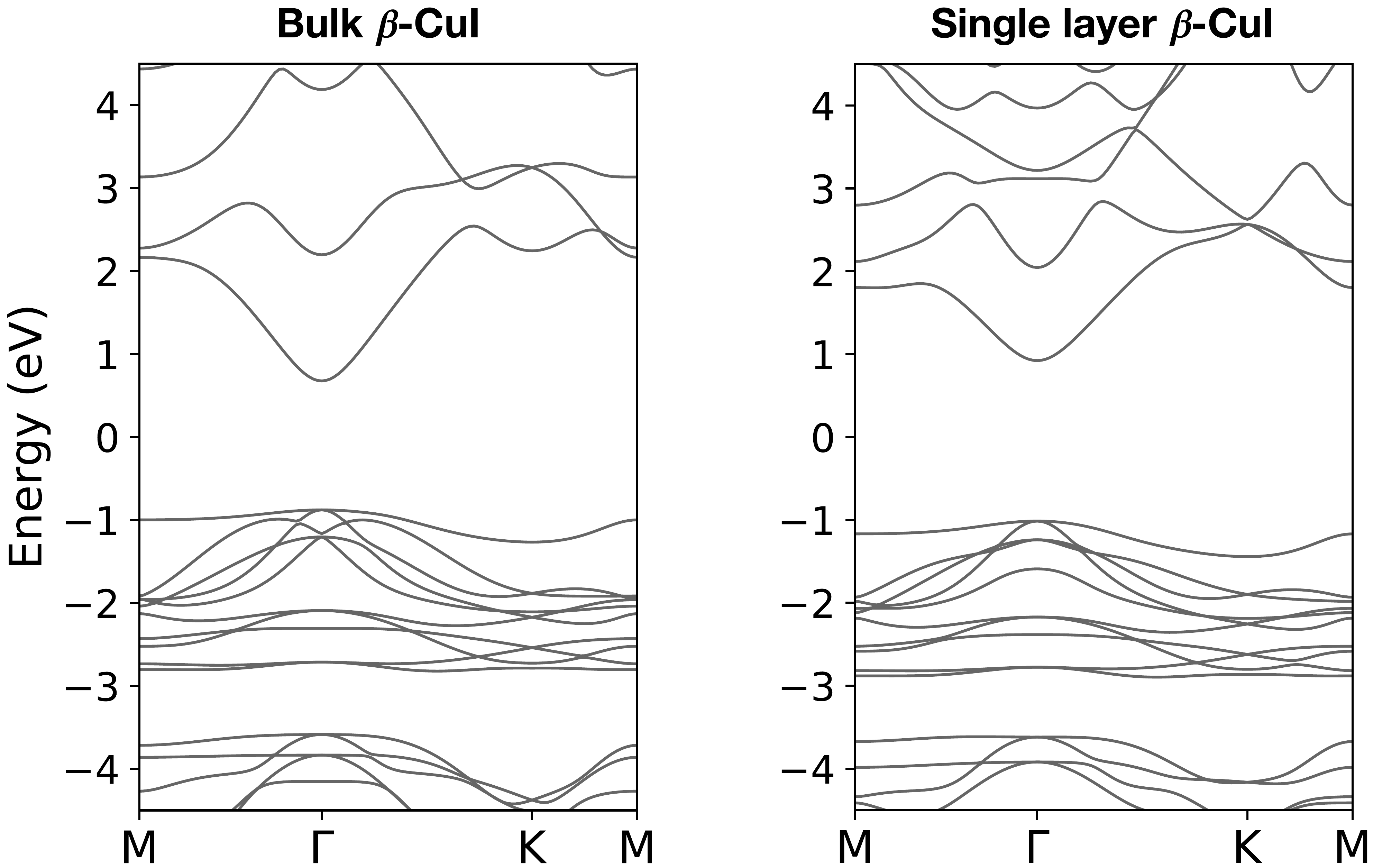}
    \caption{Band structures for the unit cell of bulk and single layer $\beta$-CuI calculated with the optPBE-vdW density functional.}
    \label{fig:bs_cui}
\end{figure}

\begin{table}[!]
\renewcommand{\arraystretch}{1.5}
\centering
 \begin{minipage}[t]{0.45\hsize}\centering
  \caption{Lattice constants and interlayer binding energy of bulk $\beta$-CuI calculated with the optPBE-vdW density functional.}
  \begin{tabular}{m{0.25\textwidth}<{\centering}m{0.25\textwidth}<{\centering}m{0.35\textwidth}<{\centering\arraybackslash}}
   \hline
   $a$,~{\AA} & $c$,~{\AA} &
   $E_\text{bind}$,~meV/{\AA}$^2$ \\
   \hline
   4.203     & 7.345     &  13.40 \\  
   \hline
  \end{tabular}
 \end{minipage}%
 \hfill
 \begin{minipage}[t]{0.45\hsize}\centering
  \caption{Band gap in eV for bulk and single layer $\beta$-CuI calculated with the optPBE-vdW and HSE06 functionals.}
  \begin{tabular}{m{0.3\textwidth}<{\centering}m{0.35\textwidth}<{\centering}m{0.3\textwidth}<{\centering\arraybackslash}}
  
      & optPBE-vdW & HSE06 \\
  \hline
  Bulk         & 1.554 & 2.683 \\
  Single layer & 1.936 & 3.172 \\  
  \hline
  \end{tabular}
 \end{minipage}
\end{table}

\begin{figure}[t!]
	\includegraphics[width=0.95\textwidth,keepaspectratio]{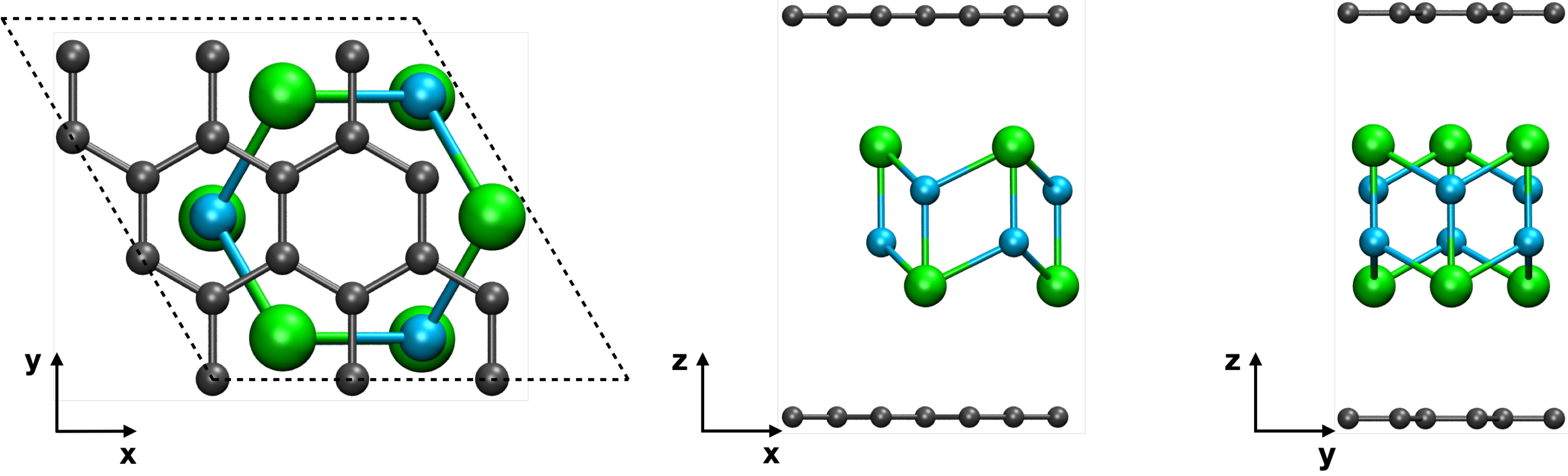}
    \caption{Hexagonal unit cell of h-CuI/graphene heterostructure.}
    \label{fig:gr_cui}
\end{figure}

\begin{figure}[!]
	\includegraphics[width=0.95\textwidth,keepaspectratio]{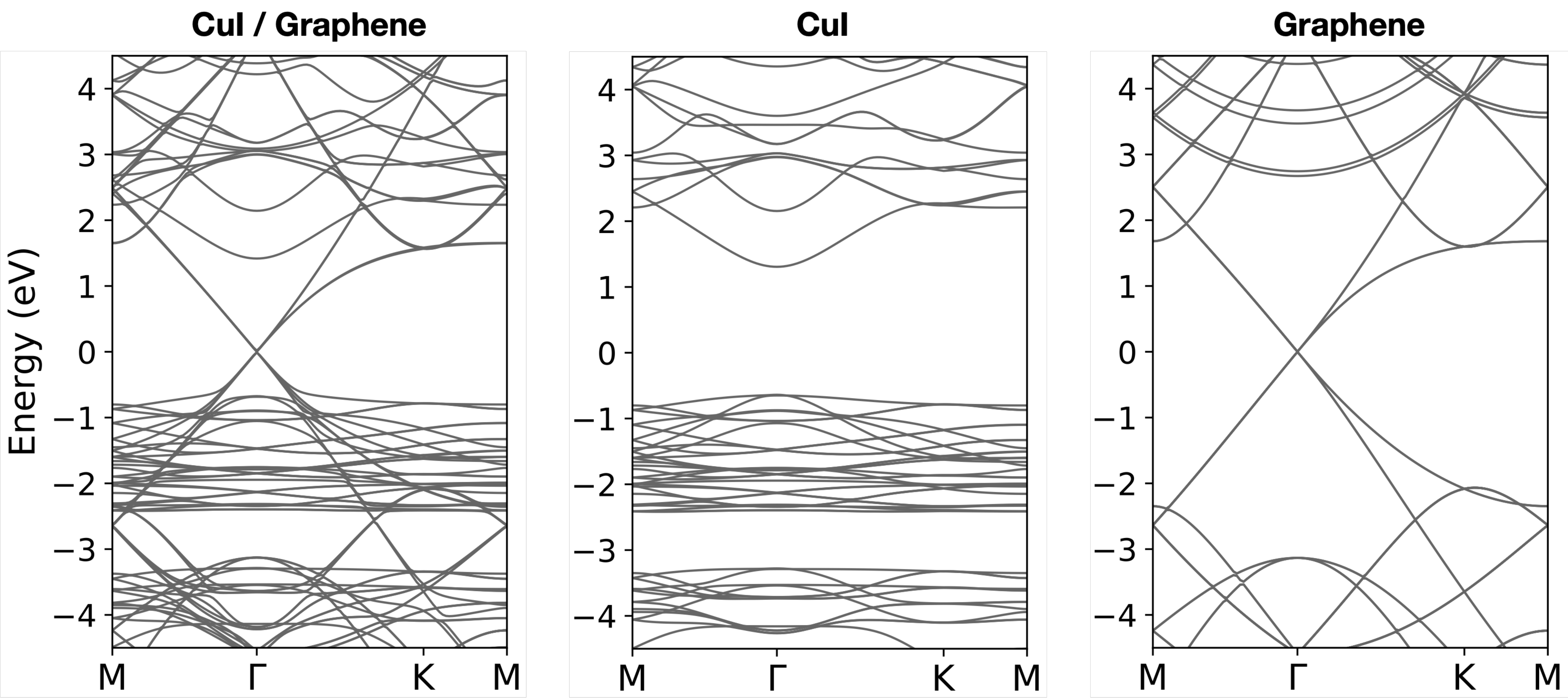}
    \caption{Band structures of h-CuI/graphene heterostructure, $2\times2$ supercell of h-CuI and $3\times3$ supercell of graphene calculated with the optPBE-vdW density functional. Zero energy corresponds to the position of the Fermi level.}
    \label{fig:bs_gr_cui}
\end{figure}

\begin{table}[!]
\caption{Distances between graphene and CuI, $d_\text{Gr-CuI}$, and binding energies per graphene layer, $E_\text{bind}$, for different stacking of graphene and CuI layers. H, B, T correspond to configurations where iodine atoms are positioned above (below) the centers of graphene hexagons, middles of the C--C bonds, and  C atoms, respectively.}
\renewcommand{\arraystretch}{1.5}
\centering
\begin{tabular}{m{0.2\textwidth}<{\centering}m{0.2\textwidth}<{\centering}m{0.2\textwidth}<{\centering\arraybackslash}}
\hline
  Position & $d_\text{Gr-CuiI}$,~{\AA} &
  $E_\text{bind}$,~meV/{\AA}$^2$ \\
\hline  
  H        & 3.666 & 14.18 \\
  B        & 3.679 & 14.05 \\
  T        & 3.691 & 14.03 \\    
\hline
\end{tabular}
\end{table}

\clearpage

\subsection{Synthesis of other 2D metal-halides within graphene films}

\begin{figure}[!]
	\includegraphics[width=1.0\textwidth,keepaspectratio]{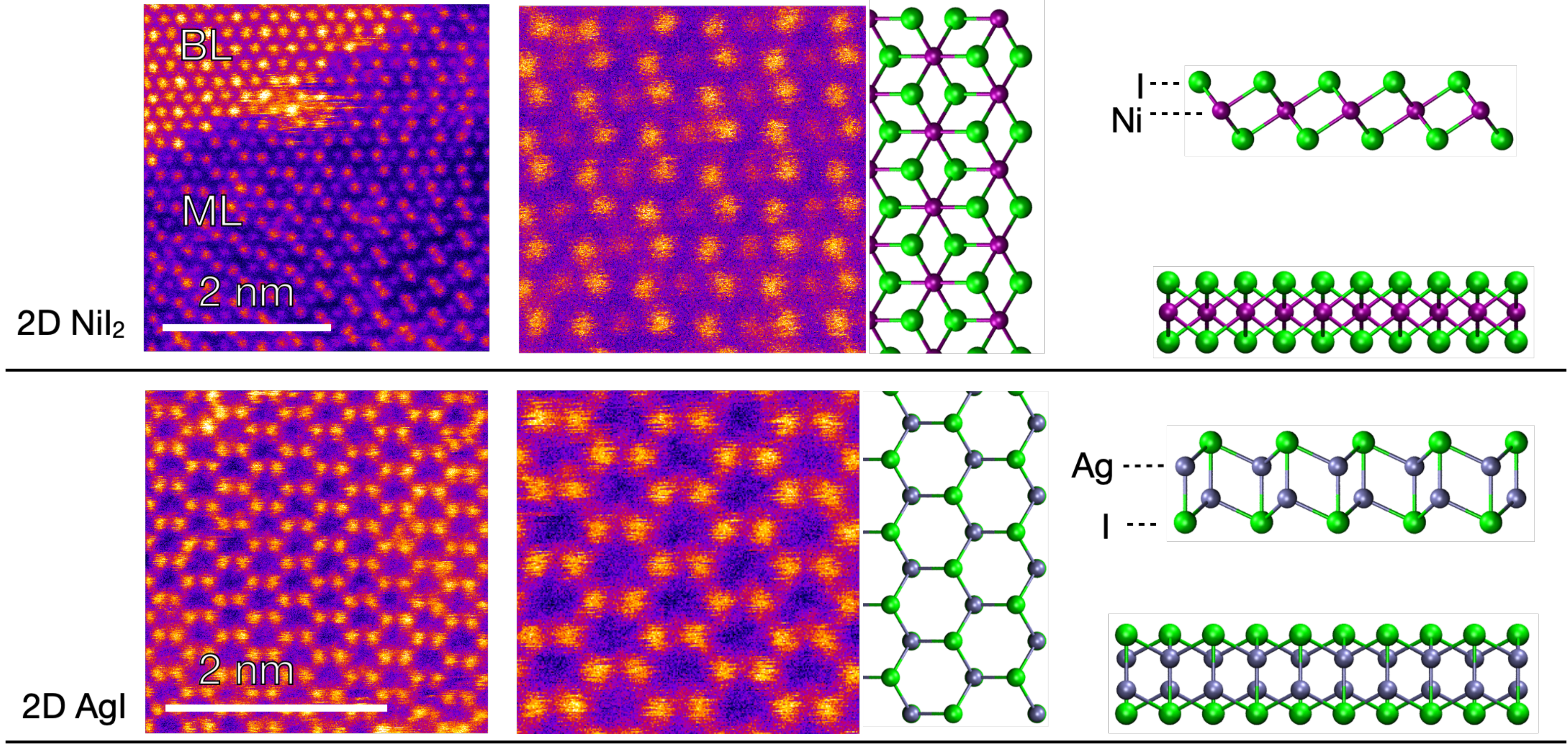}
    \caption{STEM HAADF images of the further synthesized 2D metal-halides: mono- (ML) and bi-layers (BL) NiI$_2$, and a monolayer of AgI. The images are false-coloured with ImageJ lookup table Fire.}
    \label{fig:metal_halides}
\end{figure}

\end{suppinfo}

\end{document}